\documentstyle[12pt,epsfig]{article}

\large
\newcommand{\be}{\begin{eqnarray}}
\newcommand{\ee}{\end{eqnarray}}
\newcommand\del{\partial}
\newcommand{\Dirac}{\rlap {\hspace{-0.5mm} \slash} D}

\begin{document}
\setlength{\baselineskip}{17pt}
\pagestyle{empty}
\vfill
\eject
\begin{flushright}
SUNY-NTG-98/25
\end{flushright}

\vskip 2.0cm
\centerline{\Large \bf  From chiral Random Matrix Theory}
\vskip 1cm 
\centerline{\Large\bf to chiral Perturbation Theory}
\vskip 1.2cm
\centerline{J.C. Osborn, D. Toublan and J.J.M. Verbaarschot}
\vskip 0.2cm
\centerline{\it Department of Physics and Astronomy, SUNY, 
Stony Brook, New York 11794}
\vskip 1.5cm

\centerline{\bf Abstract}
We study the spectrum of the QCD Dirac operator by means of
the valence quark mass dependence of the chiral condensate
in partially quenched Chiral Perturbation Theory (pqChPT) in the 
supersymmetric formulation of Bernard and Golterman. 
We consider valence quark masses
both in the ergodic domain ($m_v \ll E_c$) and the 
diffusive domain ($m_v \gg E_c$). These domains are separated by a mass scale
$E_c \sim F^2/\Sigma_0 L^2$ (with $F$ the pion decay constant, $\Sigma_0$ the 
chiral condensate and $L$ the size of 
the box). In the ergodic domain  the effective super-Lagrangian
reproduces the microscopic spectral density of chiral Random Matrix
Theory (chRMT). We obtain a natural explanation of 
Damgaard's relation 
between the spectral density and the finite volume partition function with 
two additional flavors.  We argue that in 
the ergodic domain the natural measure
for the superunitary integration in the pqChPT partition function is
noncompact.
We find that the tail of the two-point spectral correlation
function derived from pqChPT agrees with the chRMT  result in the 
ergodic domain.
In the diffusive domain we extend the results for the slope of the
Dirac spectrum first obtained by Smilga and Stern. 
We find that the spectral density diverges logarithmically for nonzero
topological susceptibility. We study the transition
between the ergodic and the diffusive domain and identify a range where
chRMT and pqChPT coincide. 
\vskip 0.5cm
\noindent
{\it PACS:} 11.30.Rd, 12.39.Fe, 12.38.Lg, 71.30.+h// /noindent
{\it Keywords:} QCD Dirac operator; Chiral random matrix theory; Partially
quenched chiral perturbation theory; Thouless energy; Microscopic spectral
density; Valence quark mass dependence

\vfill
\noindent

\eject
\pagestyle{plain}

\vskip1.5cm
\noindent
{\bf 1. Introduction}
\vskip 0.5cm

Spontaneous breaking of chiral symmetry is believed to be an important
property of the strong interaction. In the limit of massless quarks the
QCD Lagrangian is invariant under $U_L(N_f) \times U_R(N_f)$ transformations,
but the
vacuum state is not. The analysis of  hadron spectra and numerical
simulations \cite{DeTar,Ukawa,Smilref} 
on the lattice strongly support this assertion. 
The order parameter for  spontaneous breaking of chiral symmetry
is the chiral condensate. In the chiral limit it 
is related to the spectral density of
the QCD Dirac operator by means of the Banks-Casher formula \cite{BC},
\be
\Sigma_0 = \lim \frac {\pi \rho(0)} V,
\label{Banks-Casher}
\ee 
where $V$ is the volume of space-time and the spectral density is
defined by
\be 
\rho(\lambda)= \sum_k \delta(\lambda-\lambda_k). 
\label{eq:rho} 
\ee 
Here, the $\lambda_k$ are the eigenvalues of the Euclidean QCD Dirac operator.
They  will be called virtualities in this paper.
In (\ref{Banks-Casher}) the thermodynamic limit is taken before
the chiral limit. For broken chiral symmetry, the eigenvalues near zero are
spaced as $1/\rho(0) = \pi/\Sigma_0  V$. In order to study the behavior
of the Dirac spectrum in the approach to the thermodynamic limit it is 
therefore natural to introduce the microscopic variable \cite{SVR}
\be
u = V\lambda \Sigma_0,
\ee
and the microscopic spectral density \cite{SVR} defined by
\be
\rho_S(u) = \lim_{V\rightarrow \infty} \frac 1{V\Sigma_0} \langle
\rho(\frac u{V\Sigma_0})\rangle.
\label{rhosu}
\ee  
In general, the average $\langle \cdots \rangle$ in this equation is over the 
probability distribution of the eigenvalues.
It was conjectured in \cite{SVR} that the microscopic spectral density
is universal and is given by chiral Random Matrix Theories (chRMT) with
the global symmetries of the QCD partition function. This conjecture 
has been verified both by instanton liquid \cite{Vinst} and by lattice
QCD \cite{Tilo,many,Guhr-Wilke,Tilomass} simulations. 
It was also found from lattice QCD simulations
that eigenvalue correlations on the microscopic scale, both near zero
virtuality and in the bulk of the spectrum, are given
by Random Matrix Theory \cite{Halasz,Ma,markum,Guhr-Wilke}. 
The  foundation for these observations 
was laid by  universality studies 
\cite{Damgaard,brezin,GWu,Sener1,Seneru,andystudent,Tilodam,Senerprl} 
in which it was
shown that the microscopic spectral density and other spectral correlations 
on the  scale of  individual level spacings are not affected by
macroscopic deformations of the random matrix ensemble. 
The interpretation is that
quarks undergo a chaotic/diffusive motion in four Euclidean dimensions and
one artificial time dimension. Such picture is also suggested
by the success of microcanonical lattice QCD 
simulations \cite{microcanonical}.
According to the Bohigas conjecture \cite{bohigas} 
the correlations of the eigenvalues on the microscopic
scale are then given by Random Matrix Theory. In Ref. \cite{Stern} 
chiral symmetry breaking was also related to transport properties of
quarks in  4+1 dimensions.

Of course, the dynamics of QCD, which is not included in chRMT, should 
enter in the  Dirac spectrum. 
Therefore, we expect a scale above which the microscopic spectral density
and the spectral correlations are not given by chRMT anymore.
The existence of such scale is well known in the context of mesoscopic
physics (see\cite {HDgang,Beenreview,Montambaux} for reviews). 
It is known as the Thouless energy, which is the inverse diffusion time
of an electron through the sample \cite{Altshuler}
\be
E_c = \frac {\hbar D}{L^2},
\ee
where $D$ is the diffusion
constant and $L$ is the linear dimension of the box.
In the case of QCD, the diffusion process is that of a quark propagating
through the Yang-Mills fields in a 4+1 dimensional space time. 
A second energy scale that enters in  mesoscopic physics is $\hbar/\tau_e$,
where $\tau_e$ is the elastic diffusion time. 
Based one these two scales three different domains for
the energy difference, $\delta E$, that enters in the two-point
correlation function, can be distinguished:
the ergodic domain, the diffusive domain 
or Altshuler-Shklovskii domain and the ballistic
domain.  
For energy differences $\delta E \ll E_c$ eigenvalue 
correlations are given by RMT. 
Since for time scales larger than the diffusion time, an initially localized
wave packet explores the complete phase space, 
this regime is known as the ergodic regime. In the diffusive
 regime, defined by
 $E_c \ll \delta E\ll \hbar/\tau_e$ 
only  part of the phase space
is explored by an initially localized  wave packet, 
resulting in the disappearance of eigenvalue correlations.
In this paper we don't consider the ballistic regime with $\delta E \gg 
\hbar/\tau_e$. 
For an interpretation of the Thouless
energy in terms of the spreading width we refer to \cite{spreading,Guhr}.
As was shown in \cite{Imry,kravtsovmoriond},
the spectral two-point function can be related to
the semiclassical return probability which provides a simple intuitive
picture of  its asymptotic behavior. 
For other recent studies on this topic we refer to
\cite{Altland,Braun,Aronov,yan,kravtsov-lerner,Altland-Gefen,Chalker-kravtsov}.
What has emerged from these studies is that there
is a close relation between eigenvalue correlations and localization 
properties of the wave functions.
The interpretation of spontaneous chiral symmetry 
breaking as a delocalization transition was  
made earlier in \cite{shuryak}. By analogy with the
Kubo formula, $\Sigma_0$ plays the role of the conductivity \cite{shuryak}.

In QCD Dirac spectra the Thouless energy was identified as the scale
where the inverse pion mass is of the order of the linear size
of the box \cite{Osbornprl,Janik}. Using standard
 relations based on the approximate chiral
symmetry of the QCD partition function it has been suggested that
\cite{Osbornprl,Osborn,Janik}
\be
E_c \sim \frac {F^2}{\Sigma_0 L^2},
\ee
where $F$ is the pion decay constant.
The existence of such a scale has been observed in instanton liquid
\cite{Osbornprl,Osborn} and  lattice QCD \cite{many,Guhr-Wilke} simulations.
One of the main goals of this  paper is to  
 identify this scale by means of partially quenched
Chiral Perturbation Theory \cite{pqChPT}.

In this paper we will discuss four different partition functions: the 
Euclidean QCD partition function, the effective 
finite volume partition function 
\cite{GL,LS}, the chiral random matrix theory partition function \cite{SVR}
and the partition function for partially quenched Chiral Perturbation
Theory \cite{pqChPT}. 
Below, we briefly define 
the first three partition functions whereas the 
last one is discussed extensively in the remainder of this paper. 
At this point we only mention that pqChPT is an effective mesonic theory
which is generally believed to describe 
the low-energy limit of QCD. In addition to the usual $N_f$ sea
quarks, this theory has $k$ (in this paper $k = 1$) valence quarks and
an equal number of bosonic ghost quarks with the same mass $m_v$. Without
a source term that breaks the supersymmetry the valence
sector drops out of the partition function. In other words, the valence
quark mass only appears in the operators.

The Euclidean QCD partition function with fundamental quarks
is given by 
\be 
Z_{QCD}({\cal
M})=\int dA \; {\rm det}(\Dirac + {\cal M}) \; e^{-S_{YM}},
\label{eq:ZQCD} 
\ee 
where ${\cal M}$ is the quark mass matrix,
$\Dirac$ is the Euclidean Dirac operator and $S_{YM}$ is the
Yang-Mills action. The integral is over all gauge field configurations $A$.
The
eigenvalues of the Dirac operator are defined by 
\be 
\Dirac \psi_k= i\lambda_k \psi_k.  
\ee 
Notice that for $\lambda_k \ne 0$ the eigenfunctions occur in pairs
$\psi_k$, $\gamma_5 \psi_k$ with opposite eigenvalues.
The quark mass dependence resides in the
determinant.

For an Euclidean space-time volume with linear dimension in the range
\be
1/\Lambda \ll L \ll 1/m_\pi,
\label{range}
\ee
($\Lambda$ is a typical hadronic scale and $m_\pi$ is the pion mass)
the mass dependence of the QCD partition function is given by 
the effective finite volume partition function. For $N_f$ flavors, vacuum
angle $\theta$  and mass matrix
${\cal M}$  it is defined by \cite{GL,LS}
\be
Z^{\rm eff}_{N_f}({\cal M},\theta)
\sim \int_{U\in G/H} dU e^{V\Sigma_0 {\rm Re\,Tr}\,{\cal M} Ue^{i\theta/N_f}}.
\label{zeff}
\ee
The integral is over the Goldstone manifold associated with chiral
symmetry breaking from $G$ to $H$. For three or more colors with
fundamental fermions $G/H = SU_L(N_f) \times SU_R(N_f)/SU_A(N_f)$.
The effective finite volume
partition function in the sector of topological charge $\nu$ follows
by Fourier inversion with respect to vacuum angle $\theta$.
The partition function for $\nu = 0$ is thus given by (\ref{zeff})
with the integration over $SU(N_f)$ replaced by an integral over
$U(N_f)$. For an elaborate discussion of this partition function we
refer to \cite{LS}.

The chiral random matrix partition function
 with the global symmetries of the QCD partition function as
input is defined by {\cite{SVR,V}}
\be
Z_{N_f,\nu}^\beta(m_1,\cdots, m_{N_f}) =
\int DW \prod_{f= 1}^{N_f} \det({\rm \cal D} +m_f)
e^{-\frac{N \beta}4 \Sigma{\rm Tr}W^\dagger W },
\label{ranpart}
\ee
where
\be
{\cal D} = \left (\begin{array}{cc} 0 & iW\\
iW^\dagger & 0 \end{array} \right ),
\label{diracop}
\ee
and $W$ is a $n\times m$ matrix with $\nu = |n-m|$ and
$N= n+m$.
As is the case in QCD, we assume that the equivalent of the topological charge
$\nu$ does not exceed $\sqrt N$,
so that, to a good approximation, $n = N/2$.
Then the parameter $\Sigma$ can be identified as the chiral condensate and
$N$ as the dimensionless volume of space time (Our units are defined 
such that the density of the modes $N/V=1$). 
The matrix elements of $W$ are either real ($\beta = 1$, chiral
Gaussian Orthogonal Ensemble (chGOE)), complex
($\beta = 2$, chiral Gaussian Unitary Ensemble (chGUE)),
or quaternion real ($\beta = 4$, chiral Gaussian Symplectic Ensemble (chGSE)).
For QCD with three or more colors and quarks in the fundamental representation
the matrix elements of the Dirac operator are complex and we have $\beta = 2$.
It can be shown that in the domain (\ref{range}) the random matrix
partition function can be mapped onto the effective 
finite volume partition function
\cite{SVR}. For more discussion of the chRMT partition function we refer
to \cite{camreview}.

In this paper we study the Dirac spectrum by means of the valence quark
mass dependence of the chiral condensate \cite{Christ,vPLB,Trento}
\be
\Sigma(m_v) = \frac 1V \sum_k \left \langle \frac 1{m_v+ i \lambda_k} 
\right \rangle,
\label{discval}
\ee
where $\langle \cdots \rangle$ denotes an average with respect to
the distribution of the eigenvalues.
The spectral density follows from the discontinuity across the 
imaginary axis,
\be
\left .{\rm Disc}\right |_{m_v = i\lambda}\Sigma(m_v) 
= \lim_{\epsilon \rightarrow 0}
\Sigma(i\lambda+\epsilon) - \Sigma(i\lambda-\epsilon) = 2\pi \sum_k 
\langle \delta(\lambda +\lambda_k)\rangle 
= 2\pi \langle \rho(\lambda)\rangle.
\label{spectdisc}\nonumber \\
\ee
The two point spectral correlation function is given by
\be
\langle \rho(\lambda) \rho(\lambda') \rangle
= \frac 1{4\pi^2}\left . {\rm Disc}
\right |_{m_v = i\lambda, m_{v'}=i\lambda'}  
  \sum_{k,l}\left\langle
 \frac 1{i\lambda_k +m_v}\frac 1{i\lambda_l +m_{v'}}\right\rangle.
\ee
On the other hand, the valence quark mass dependence
can be calculated from the partially quenched partition function. By 
differentiating with respect to the valence quark masses we obtain in the
parametrization discussed in section 2,
\be
\langle \rho(\lambda) \rho(\lambda') \rangle^C =
\frac 1{8\pi^2} \left . {\rm Disc}
\right |_{m_v = i\lambda, m_{v'}=i\lambda'} \frac {\Sigma^2_0}{F_\pi^4}
\int d^4 x \int d^4 y \langle  
\xi_{vv'}(x)\xi_{vv'}(x)\xi_{vv'}(y)\xi_{vv'}(y)\rangle_{pq}^C,
\nonumber \\
\ee
where the superscript $C$ refers to the connected correlation function 
which is defined as $\langle \rho(\lambda_1) \rho(\lambda_2) \rangle^C=
\langle \rho(\lambda_1) \rho(\lambda_2) \rangle-
\langle \rho(\lambda_1) \rangle \langle \rho(\lambda_2) \rangle$. The notation
$\langle \cdots \rangle_{pq}$ denotes averaging with respect to the
partially quenched partition function.  The meson fields with quark content
corresponding to a mass $M=(m_v+m_{v'})\Sigma_0/F^2$ are denoted by $\xi_{vv'}$.
The expectation value of the product of the meson fields
 can be expressed
in terms of the meson propagator in momentum space. Taking into account
 the  combinatorial factors and  performing 
the integrals over $x$ and $y$ and one momentum integral this results in
\be
\langle \rho(\lambda) \rho(\lambda') \rangle^C =
\frac 1{4\pi^2}\frac {\Sigma^2_0}{F^4}\left . {\rm Disc}
\right |_{m_v = i\lambda, m_{v'}=i\lambda'}\sum_q\frac 1{(q^2 +M^2)^2}.
\ee
We have expressed the two-point correlation function as a sum over momentum 
modes. Such relations are well known from the theory of disordered systems
(see for example \cite{Altland,Imry,kravtsovmoriond}). 
The  zero momentum modes reproduce the asymptotic limit of the random
matrix result for the two-point correlation function. 
This can be seen by evaluating
the discontinuities after substitution of the Gell-Mann-Oakes-Renner relation
for the meson mass. For $|\lambda-\lambda'| \ll \lambda$ but much larger than
the average level spacing we obtain
\be
\langle \rho(\lambda) \rho(\lambda') \rangle^C \sim
-\frac 1{2\pi^2} \frac 1{(\lambda-\lambda')^2}.
\ee
which is the well-known random matrix result.
The full correlation function can only be obtained by means of 
an exact integration over the saddle-point manifold \cite{Andreev}. 
In this paper we will not perform this task for
the two-point function but instead focus on the one-point function.
Because of the $U_A(1)$ symmetry, the microscopic spectral density already shows
a nontrivial structure for eigenvalues close to zero as has been
observed in lattice QCD simulations \cite{Tilo,many}. The discussion of
the two-point function will appear in a future publication \cite{Toublan}.
 
The valence quark mass dependence of the chiral condensate was  
calculated in partially quenched chiral perturbation theory
by Golterman and Leung \cite{GolLeung} by means of a supersymmetric 
method introduced by Bernard and Golterman \cite{Morel,pqChPT}. We use this  
result to calculate  the
spectral density in the perturbative domain.
Among others we reproduce the 
slope of the spectral density first derived by Smilga and 
Stern \cite{Smilga-Stern}.  When, for virtualities below the 
Thouless energy, the zero modes are treated exactly, we reproduce
the random matrix result for the microscopic spectral density. In fact,
we show that the partially quenched supersymmetric partition function
can be written as an ordinary effective finite volume
 partition function with two more flavors which are the equivalent
of one valence quark flavor and its supersymmetric partner. 
In this way we obtain a natural
explanation of a  result first found by Damgaard
\cite{Dampart}.

The organization of this paper is as follows.
Partially quenched chiral perturbation theory is introduced in section 2. In
section 3 we discuss its different domains of validity. The choice of the
superunitary measure is motivated in section 4. In sections 5  and 6
the partially quenched partition function is studied in the domain where
it is dominated by the zero modes. The valence
 quark mass dependence in the quenched limit is calculated in
section 5, and, in section 6,  we show
 that the microscopic spectral density that follows from the 
valence quark mass dependence of the chiral 
condensate coincides with the result from chRMT.  
In section 7 we calculate the valence quark 
mass dependence of the condensate by
means on pqChPT. The transition from chRMT to pqChPT is discussed in section 8.
In section 9 we evaluate the spectral density in the diffusive domain.
Concluding remarks are
made in section 10. An explicit calculation of the spectral density for
$N_f=1$ is presented in Appendix A and a short derivation of a Berezinian
that occurs in the diagonalization of superunitary transformations is
given in Appendix B.

\vskip1.5cm
\noindent
{\bf 2. Partially Quenched Chiral Perturbation Theory}
\vskip 0.5cm

Partially Quenched Chiral Perturbation Theory (pqChPT) is an effective
theory describing partially quenched QCD at low
energies~\cite{pqChPT,Sharpe}. 
This is a theory with both sea quarks, appearing in
the fermion determinant, and valence quarks which only appear in the operators.
Partially quenched chiral perturbation theory interpolates
between standard (unquenched)
Chiral  Perturbation Theory (ChPT)~\cite{GaL} and completely
quenched Chiral Perturbation Theory (qChPT)~\cite{qChPT,ColPal}. 
In this section
we give a brief overview of partially quenched chiral perturbation theory
in the formulation of Bernard and Golterman. For more details we refer to
the original literature \cite{pqChPT}. 

We consider a  QCD-like theory with one valence quark and its
supersymmetric partner, both with mass $m_v$, 
and $N_f$ unquenched quarks (sea quarks) with  mass $m_s$. 
Then the fermion determinant corresponding to the valence
quark is cancelled by the contribution from a bosonic ghost quark with the same 
mass \cite{Morel}. The full chiral symmetry of
the theory is thus the direct product of the supergroups (also called 
graded groups)
$$\mbox{$G=U_L(N_f+1|1) \otimes U_R(N_f+1|1)$},$$ 
where one of the $U(1)$ groups is broken by the anomaly.
The
partially quenched QCD Lagrangian  at a given order
in the momentum can be approximated by an  effective Lagrangian for the  field
\mbox{$U \in SU(N_f+1|1) \oslash U(1)$} 
which transforms linearly under the chiral group.  
(Here, $\oslash$ denotes a semi-direct product.)
In terms of the Goldstone fields $U$ is parametrized
as
\be
 U={\rm exp} (i\sqrt 2 \Phi/F), 
\ee 
where $\Phi$ is a $(N_f+2) \times (N_f+2)$ super-Hermitian matrix field: 
\be 
\Phi=\left(
\begin{array}{ll}
\phi & \chi^\dagger \\ \chi & \tilde{\phi}
\end{array}
\right)  
\ee 
and $F$ is the pion decay constant.
Here, $\phi$ is the $(N_f+1) \times (N_f+1)$ matrix containing the
ordinary mesons made of quarks and antiquarks, $\tilde{\phi}$ is the 
meson with two ghost quarks, and $\chi$, which is  a vector of length
$N_f+1$,  represents the  
fermionic mesons consisting of a ghost-quark and an ordinary
anti-quark. We choose a diagonal mass matrix with degenerate sea quark masses
resulting in the  $(N_f+2) \times (N_f+2)$ quark-mass matrix  
\be 
\hat{\cal M}={\rm diag} (m_v,\underbrace{m_s, \dots,
m_s}_{N_f},m_v).  
\ee
The corresponding pion masses will be denoted by 
\be
M_{ij}^2 = (m_i+m_j)\Sigma_0/F^2\quad {\rm for} \quad i,j = s,v.
\ee     

Expanding the Euclidean effective Lagrangian in terms of the 
momenta and the quark masses we obtain to lowest order \cite{pqChPT,GolLeung}
\be 
{\cal L}_{\rm
eff}=\frac{F^2}{4} \; {\rm Str} (\partial_\mu U \; \partial_\mu
U^\dagger)-\frac{\Sigma_0}{2} \; {\rm Str} (\hat{\cal M} U+\hat{\cal
M} U^\dagger)+\frac{m_0^2}{6} \; \Phi_0^2+\frac{\alpha}{6} \;
\partial_\mu \Phi_0 \; \partial_\mu \Phi_0,
\label{superL}
\ee 
where we have added a mass term and a kinetic term for the super-$\eta'$
flavor-singlet field $\Phi_0={\rm Str} (\Phi)$ with parameters 
$m_0^2$ and $\alpha$ (for the definition of the supertrace, Str,  
(also called graded trace) 
we refer to \cite{VWZrep}). 
In the next section we will show that the term
containing $\alpha$ is sub-leading in the chiral expansion. 
The quark condensate in the chiral limit is denoted by $\Sigma_0$.
The partition function corresponding to (\ref{superL}) is defined by
\be
Z_{pq} = \int d U e^{-\int d^4 x {\cal L}_{\rm eff}}
\label{superpart}
\ee
and will be studied for different ranges of the
quark masses.

A physical interpretation of the zero momentum component
of the flavor singlet mass term can be obtained
by linearizing the square according to
\be
e^{-V\frac {m_0^2}6 \Phi_0^2} = \frac 1{\sqrt{2\pi\langle \nu^2\rangle}} 
\int d\nu \,
{\rm Sdet}^\nu (U)\,  e^{-\frac {\nu^2}{2\langle \nu^2\rangle}},
\ee
where
\be
\langle \nu^2\rangle = \frac {F^2 m_0^2 V}6 
\label{WV}
\ee  
is the familiar topological susceptibility. For the definition of the
superdeterminant, Sdet,  (also called graded determinant) 
we refer to \cite{VWZrep}.
This result allows us to interpret $m_0^2$  as a topological contribution to
the singlet mass. The partition function with zero total topological charge
is obtained by  multiplying the Goldstone fields with the $U_A(1)$ phase and
extending the integration manifold to include this contribution (see 
\cite{LS}).  
In the zero momentum sector, the partition function for $\nu = 0$
is thus given by (\ref{superpart}) and
the Lagrangian (\ref{superL}) without the topological 
mass term $m_0^2 \Phi_0^2/6$
but with the integration domain extended to $U(N_f+1|1)$.
This partition function will be discussed in sections 5 and 6. 
Notice that the microscopic spectral density depends on the 
global topological charge. On the other hand, the spectral density at a 
macroscopic distance from zero 
is not sensitive to the $global$ topological charge.

\vskip1.5cm
\noindent
{\bf 3. Domains of Validity}
\vskip 0.5 cm

For standard chiral perturbation theory the different domains of its validity 
were discussed in detail by Gasser and Leutwyler \cite{GL}. 
Since the discussion for 
partially quenched chiral perturbation theory is quite analogous
we refer to this paper for additional arguments.

With the basic scale given by $\epsilon \equiv 1/L \ll \Lambda_{QCD}$, 
we consider five
 different mass scales: the smallest eigenvalue of the
Dirac operator $\lambda_{\rm min}\sim \epsilon^4$, the valence quark 
mass $m_v\sim \epsilon^\alpha$, the sea quark
mass $m_s \sim \epsilon^\beta$, the pion momenta $p_\mu \sim 1/L \sim 
\epsilon$ and the $\eta'$ mass $m_{\eta '} \sim \Lambda_{\rm QCD} \sim 1$.

The integral over $U$ in (\ref{superpart}) 
can be split into an integral over the zero momentum
modes, $U_0$, and the nonzero momentum modes, $\xi(x)$, i.e.
\be
U = U_0 e^{i\xi(x)}.
\ee
The two different types of modes are mixed by the mass terms and by the 
singlet interaction. Keeping only
terms up to second order in $\xi$, our Lagrangian can be written as
\be 
\int d^4 x {\cal L}_{\rm eff}=&-&\frac{\Sigma_0 }{2} \; 
{\rm Str}\left [  \hat{\cal M} (U_0+U^\dagger_0)(V-\frac 12 
\int d^4 x \xi^2(x) )\right ]
\nonumber \\
&+& \; \int d^4 x 
\left (\frac{F^2}{4} \partial_\mu \xi_a \; \partial_\mu
\xi_a+\frac{m_0^2}{6} \; \Phi_0^2+\frac{\alpha}{6} \;
\partial_\mu \Phi_0 \; \partial_\mu \Phi_0 \right ). 
\label{superLdomain}
\ee 
The nonzero mode fields are $O(\epsilon)$ whereas the zero mode fields
are $O(\epsilon^{(4-\alpha)/2})$. The term that mixes 
both types of modes is  of  order $\epsilon^2$ independent of the
value of $\alpha$  (for $1 \le \alpha<4$). This implies that the 
mixing term is always subleading. Since the fluctuations of the singlet
field are of $O(\epsilon^2)$, the kinetic term of the singlet field is 
subleading as 
well.

Since the mass term of the nonzero modes is of order $\epsilon^{\alpha- 2}$ 
the integral over the nonzero modes does not contribute to the mass
dependence of the partition function for $\alpha > 2$. Therefore, for
$\alpha < 2$ (or $m < 1/\Lambda_{\rm QCD} L^2$)
the mass dependence is completely determined by the zero modes. 
The crossover-point can be determined more accurately
by considering the mass dependence of the propagator of the nonzero modes.
We will return to this point in section 8.

In the chiral limit, $m_s \rightarrow 0$, the integrals over the zero momentum
modes become perturbative for $m_v \gg \lambda_{min}$, whereas for $m_v \ll
1/\Lambda_{\rm QCD} L^2$ the mass dependence of the partition function is
completely determined by the zero momentum modes.
In the domain
\be
\lambda_{\rm min} \ll m_v \ll 1/\Lambda_{\rm QCD} L^2
\label{valdomain}
\ee 
the valence quark mass dependence 
of the chiral condensate can thus be determined both from
the integral over the zero momentum modes 
and from a perturbative calculation.
In section 8 it will be shown that this consistency condition is satisfied.

Actually, because of the mixing by the singlet mass term,
 the above discussion has to be refined by considering modes that
are eigenvectors of the mass matrix.
This is essential for the quenched limit. In this case
the relevant mesonic mass matrix in a quark basis is given by \cite{qChPT}
\be
\left ( \begin{array}{cc}
\frac {m_0^2}3 + M^2_{vv}& -\frac {m_0^2}3 \\
        -\frac {m_0^2}3 & \frac {m_0^2}3 -M^2_{vv}   
\end{array} \right ).
\ee
To lowest order in $M_{vv}$ the eigenvalues of this matrix are given by
$2m_0^2/3 + M_{vv}^4/(2m_0^2/3)$ and $-M_{vv}^4/(2m_0^2/3)$. The zero
momentum mode corresponding to the second eigenvalue only becomes 
perturbative for $m_v \gg \epsilon^2$, where also the nonzero modes 
contribute to the mass dependence. We conclude that in the 
quenched case there is no common domain of validity.

Let us contrast this with the mesonic mass matrix for $N_f = 1$ which is given 
by
\be
\left ( \begin{array}{ccc}
\frac {m_0^2}3 + M^2_{vv}& \frac {m_0^2}3&-\frac {m_0^2}3 \\
        \frac {m_0^2}3& \frac {m_0^2}3+M^2_{ss} & -\frac {m_0^2}3\\   
        -\frac {m_0^2}3 &-\frac {m_0^2}3& \frac {m_0^2}3 -M^2_{vv} 
\end{array} \right ).
\ee
In the chiral limit ($M_{ss}\rightarrow 0$) the eigenvalues up to order
$M_{vv}^4$ are given by $\lambda_ 1 = m_0^2 + 2M_{vv}^4/3$,
$\lambda_2 = -M^2_{vv}/\sqrt 3 - M_{vv}^4/3$
and
$\lambda_3 = M^2_{vv}/\sqrt 3  - M_{vv}^4/3$.
In this case, the zero 
momentum modes corresponding to the smallest eigenvalue become perturbative
for $m_v \gg \lambda_{min}$ and  a common domain of validity can be 
identified.
For $N_f$ massless sea quark flavors, $N_f-1$ eigenvalues are zero 
in the diagonal basis. 
However, mesons with two sea quarks do not enter in the one-loop calculation
of the valence quark mass dependence of the chiral condensate.
The smallest non-zero eigenvalue in the valence sector is again
of order $M_{vv}^2$. Therefore, as for one flavor, a common domain of 
validity can be found. Of course, if also the sea quark masses are in the
range (\ref{valdomain}) a common domain of validity for 
the dependence of the chiral condensate on the sea quark masses
can be found as well. 

\vskip1.5cm
\noindent
{\bf 4. Choice of Measure}
\vskip 0.5cm

For the group integration  in (\ref{superpart}), we have two 
different possible choices for 
the superunitary integration measure:  a compact one
which was used in \cite{qChPT,pqChPT},
and  a noncompact one which was 
introduced in \cite{Andreev,zirnall,Sener1} in the context of 
chiral Random Matrix Theory. 
It was argued in \cite{Sener1} that a noncompact measure 
reproduces the invariance of the bosonic degrees of freedom under 
noncompact $U_A(1)$ transformations and therefore should be
the appropriate measure for the quenched theory. 
We stress that for perturbative
calculations, as in \cite{qChPT,pqChPT}, both measures lead to the 
correct result.

For the purely quenched theory a decomposition of
the compact measure in terms of the eigenvalues of the 
superunitary matrix $U$
can be written down easily. 
The compact measure for the
group $U(1|1)$ can thus be written as
\be
D U \sim \frac {d\theta d\phi}{|e^{i\theta} - e^{i\phi}|^2} D U',
\label{compact}
\ee
in the parametrization
\be
U =  U' \Lambda U'^{-1},
\label{decompose}
\ee
with $\Lambda$ a diagonal matrix with eigenvalues $\exp(i\theta)$ and 
$\exp(i\phi)$. This decomposition can be easily generalized to an arbitrary 
superunitary group. The generalization of the Berezinian appearing in
(\ref{compact}) is given in Appendix B. 

The measure for  the noncompact version of $U(1|1)$ 
is obtained by analytic continuation to
imaginary angle $\theta \rightarrow -is$ with 
$s \in (-\infty; \infty)$. This  results in the measure
\be
D U \sim \frac {ds d\phi}{(e^{s} - e^{i\phi})(e^{-s} - e^{-i\phi})} D U',
\label{noncompact}
\ee
which coincides with the measure that was obtained in \cite{Andreev,zirnall}
after a lengthy calculation. 
We will show in section 5 
that the quenched partition function with this noncompact measure
reproduces the quenched valence quark mass dependence of 
chiral random matrix theory.

The generalization of this measure to an
arbitrary superunitary group appears to be more involved than 
this  naive analytic continuation.  
However, it can be shown that the integration
contours contributing to the $discontinuity$ 
of the valence quark mass dependence
of the chiral condensate  
across the imaginary axis can be deformed such that only an integral
over a compact domain remains \cite{Andreev,zirnall,Sener1}. 
In terms of the partition
function this corresponds to replacing the noncompact measure by a compact
one. This will be the starting point of the calculation of the microscopic
spectral density discussed in section 6. 
The  calculation of the valence quark mass dependence
of the chiral condensate for arbitrary $N_f$ using a noncompact measure
will be discussed elsewhere \cite{damgandus}.

\vfill
\vskip1.5cm
\noindent
{\bf 5. Valence Quark Mass Dependence of the Chiral Condensate
in the quenched limit}
\vskip 0.5cm

In this section we will show that in the quenched case
the noncompact measure (\ref{noncompact}) leads to the correct result 
for the valence quark mass dependence of the chiral condensate. 
As argued at the end of section 2, the partially quenched partition function
in the sector of zero topological charge is obtained by 
extending the integration over $U$ in (\ref{superpart}) to $U(N_f+1|1)$
and ignoring the singlet channel mass term in (\ref{superL}). In the quenched
limit the integration is thus over $U(1|1)$.
With the measure (\ref{noncompact}) the  sector of the zero momentum modes of 
the partition function (\ref{superpart})
in the quenched limit can be written as
\be
Z(J) =\int \frac{dU d\theta ds}{(e^{s} - e^{i\theta})
(e^{-s} - e^{-i\theta})}
\exp\left[ \frac{\Sigma_0 V}2 {\rm Str}
\left ( \begin{array}{cc} m+J & 0\\ 0 & m-J \end{array} \right )
U \left ( \begin{array}{cc} \cos \theta & 0\\ 0 & \cosh s \end{array} \right )
U^{-1}\right ].\nonumber \\
\ee
The normalization of this partition function, $Z(0)$, follows from Wegner's
theorem for superinvariant functions \cite{Wegner}. 
The integral over $U$ can be performed by means of the supersymmetric version
of the Itzykson-Zuber integral \cite{Guhr91,Alfaro,GW}.
After differentiation with respect to $J$ the
result for the valence quark mass dependence of the chiral condensate is
\be
\Sigma(x)  = \frac{\Sigma_0}\pi  \int d\theta ds \frac {\cos \theta - \cosh s}
{(e^{s} - e^{i\theta})(e^{-s} - e^{-i\theta})} e^{x(\cos \theta - \cosh s)},
\ee
where we have introduced the notation $x=mV\Sigma_0$.
This integral can be evaluated most conveniently after differentiation with
respect to $x$
\be
\del_x \Sigma(x) = \frac {\Sigma_0}{4 \pi} \int d\theta ds 
(1- e^{-i\theta -s})(1- e^{i\theta +s})e^{x(\cos \theta - \cosh s)},
\label{delsig}
\ee
where the integrand has been simplified by means of the identity
\be
\frac{\cos \theta - \cosh s}{e^{i\theta} - e^s} = \frac 12 (1- e^{-s}
e^{-i\theta} ).
\label{identity}
\ee
The integrals appearing in (\ref{delsig}) factorize and can be expressed
in terms of the modified Bessel functions $K_0$ and $I_0$,
\be
\del_x \Sigma(x) = \Sigma_0[I_0(x)K_0(x) - I_1(x) K_1(x)].
\ee
The r.h.s. of this equation can be rewritten as
$ \del_x[x I_0(x) K_0(x) + x I_1(x) K_1(x)] $. This equation can be
integrated trivially with an integration constant that follows from the
asymptotic limit $\Sigma(x) \rightarrow \Sigma_0$ for $x \rightarrow \infty$.
The final result is
\be
\Sigma(x)=  \Sigma_0 x[ I_0(x) K_0(x) +  I_1(x) K_1(x)],
\label{valenceres}
\ee
which agrees with the valence quark mass dependence obtained from chiral
Random Matrix Theory \cite{vPLB}. In its domain of validity,
this result reproduces the valence quark mass dependence
found in lattice QCD
\cite{vPLB,Christ}.

\vskip 1.5cm
\noindent
{\bf 6. Microscopic Spectral Density in the Ergodic Domain}
\vskip 0.5cm

In this section we consider  the zero momentum component of the
partially quenched partition function for zero total topological charge.
In \cite{Andreev,Sener1} it was shown that the noncompact integral
contributing to the imaginary part of the chRMT 
partition function can be rewritten as an integral over $[0, 2\pi]$. 
The spectral density can then be derived from a compactified partition
function.
By a direct translation of these arguments to the partition function of pqChPT,
we obtain for the spectral density
\be
\langle \rho(\lambda) \rangle = \frac 1{2\pi}{\rm Im}  
\frac 1{Z^0_{pq}(J=0)}
\left . \frac {\del Z^0_{pq}}{\del J}\right |_{m_v = i\lambda, J= 0},
\label{spectc}
\ee
where $Z^0_{pq}(J)$ is the zero momentum 
pqChPT partition function with the {\it compact} 
measure defined by
\be
Z_{pq}^0(M) = \int d U e^{ \frac {\Sigma_0}2{\rm Str} ( M U+ M U^\dagger)}.
\ee
Here, the integral is over $U(N_f+1|1)$ corresponding to 
the sector of zero topological charge.
The mass-matrix $M$ is diagonal with matrix elements $(m_v+J, m_s, \cdots,
m_s, m_v-J)$. 

In a diagonal representation of the of superunitary matrices $U$, 
$Z^0_{pq}$ for $N_f$ flavors can be written as
\be
Z^0_{pq}(M) = \int  \prod d\theta_k dU |B(\Lambda)|^2 e^{\frac{\Sigma_0 V}2 
{\rm Str} M U (\Lambda 
+\Lambda^*) U^{-1}}.
\ee  
Here, $|B(\Lambda)|^2$ is the Berezinian for the transformation to
eigenvalues and eigenvectors as new integration variables (see Appendix B). 
The quantity 
$B(\Lambda)$ is given by
\be
B(\Lambda) = \frac{\prod_{k<l}(\Lambda_k^b - \Lambda_l^b)
\prod_{k<l}(\Lambda_k^f - \Lambda_l^f)}
{\prod_{k,l}(\Lambda_k^b - \Lambda_l^f)},
\ee
where the bosonic and fermionic eigenvalues of 
$U$ are denoted by $\Lambda_k^b$ and $\Lambda_k^f$, respectively. 
In our case we have $N_f +1$ fermionic eigenvalues 
and one bosonic eigenvalue.
In the quenched case we use the 
supersymmetric normalization with $Z^0_{pq}(J=0) = 1$ whereas in the
partially quenched case $Z^0_{pq}(J=0) = Z^{\rm eff}_{N_f}$ (defined  in 
eq. (\ref{zeff})).
Here, and below the constants multiplying the partition function
are absorbed in the integration measure. 
The integration over $U\in U(N_f+1|1)$ 
can be performed by means of a supersymmetric version
of the Itzykson-Zuber integral \cite{Guhr91,Alfaro,GW}. For $J\ne 0$
this results in
\be
Z^0_{pq}(M)= \int \prod_k d\theta_k \frac{|B(\Lambda)|^2}
{B(\cos \theta_k) B(M_k V \Sigma_0)}
e^{-M_{N_f+1}V \Sigma_0\cos \theta_{N_f+1}}
\det_{0\le k,l\le N_f} e^{M_k V \Sigma_0 \cos \theta_l}. 
\label{zcgen}
\ee
For $J= 0$, the Grassmannian components of the integration paths lead to
additional contributions. These so called Efetov-Wegner 
\cite{Efetov,Wegner}
terms only enter 
in the normalization of the partition function 
(see \cite{zirnander,zirnall,Guhr,Rothstein}
for more discussion of this point).  
 
We evaluate this integral in the chiral limit, $m_s \rightarrow 0$, for a fixed
value of the valence quark mass. In this limit the ratio of the determinant
and $B(M_kV\Sigma_0)$ is finite. 
In Appendix A, we evaluate the integral for
$N_f = 1$ and indicate how it can be calculated for arbitrary $N_f$.
In the remainder of this section we show that this generating 
function can be expressed as an ordinary 
effective 
finite volume partition function (see (\ref{zeff})) with $N_f + 2$ flavors.

For arbitrary $N_f$ the partition function (\ref{zcgen}) can be written as
\be
Z^0_{pq}(M) = {\cal N}\int \prod_k d \theta_k 
\prod_{0 \le k<l \le N_f} \frac{\omega(\theta_k, \theta_l)}{(\mu_k-\mu_l)}
\prod_{0 \le k \le N_f} \frac{(\mu_k-\mu_{N_f+1})}
{\omega(\theta_k, \theta_{N_f+1})}
e^{-\mu_{N_f+1} \cos \theta_{N_f+1}}
\det_{0 \le k,l \le N_f} e^{\mu_k \cos \theta_l}, 
\nonumber \\
\ee
with $\mu_0=(m_v+J)V\Sigma_0$,  $\mu_{N_f+1}=(m_v -J)V\Sigma_0$   and 
$\mu_k =m_s V \Sigma_0,\, k = 1, \cdots N_f$. For equal sea quark 
masses the ratio of the last factor and $\prod_{k<l}(\mu_k-\mu_l)$
has to be understood as a limit. The normalization factor that enters
in the partition function is denoted by ${\cal N}$.
The 
function
$\omega(\theta,\phi)$ defined by   
\be
\omega(\theta,\phi) = \frac{|e^{i \theta}-e^{i \phi}|^2}{\cos \theta - \cos 
\phi},
\ee
has the important property that
\be
\omega(\theta,\phi) = \frac 4{\omega(\theta,-\phi)} = 4
\frac{\cos \theta - \cos \phi}{|e^{i \theta}-e^{-i \phi}|^2}.
\ee
This allows us to bring all factors $\omega(\theta_k, \theta_l)$
into the numerator. Slightly rearranging the different factors and changing
the sign of $\mu_{N_f +1}$ we can write the partition function in 
the more symmetric form
\be
Z^0_{pq}(\tilde M) = {\cal N} \prod_{0 \le k \le N_f} (\mu_k^2-\mu_{N_f+1}^2) 
\int \prod_k d\theta_k 
\prod_{0 \le k<l \le N_f+1} \frac{4\omega(\theta_k, \theta_l)}{(\mu_k-\mu_l)}
\frac{\det_{0 \le k,l \le N_f+1} e^{\mu_k \cos \theta_l}}{N_f+2},
\nonumber\\
\ee
where the diagonal mass matrix $\tilde M$ is defined by 
$\tilde M = (m_v+J, m_s, \cdots, 
m_s, -m_v+J)$. In this integral we recognize the factors that occur in
the ordinary Itzykson-Zuber integral. It can thus be rewritten as
\be
Z^0_{pq}(\tilde M) ={\cal N}' \prod_{0 \le k \le N_f} (\mu_k^2- \mu_{N_f+1}^2) 
\int dU \prod_k d\theta_k |\Delta (e^{i\theta_k})|^2
e^{\frac {\Sigma_0 V}2 {\rm Tr} \tilde M U \cos(\theta_k) U^{-1}},
\ee
where the $U$-integral is over $U(N_f+2)$.
Up to a constant factor this is exactly the effective finite volume
partition function for $N_f + 2$ flavors. In the microscopic limit with
quark masses of order $\sim 1/V\Sigma_0$, this partition function 
is equivalent to the random matrix partition
function with $N_f +2$ flavors. For masses in the microscopic domain 
we thus have
\be
Z^0_{pq}(\tilde M) = {\cal N}'' \prod_{0 \le k \le N_f} (\mu_k^2-\mu_{N_f+1}^2) 
Z_{RMT}^{(N_f+2)}(m_v+J, m_1,...,m_{N_f},-m_v+J),
\label{zcinbetween}
\ee
where $Z_{RMT}^{(N_f+2)}$ is the chRMT partition function as defined 
in (\ref{ranpart}) for $N_f +2$ flavors, zero topological charge 
and $\beta =2$. 
Since the RMT partition function is even in all masses 
(\ref{zcinbetween}) can be rewritten as
\be
Z^0_{pq}(\tilde M) = {\cal N}''\prod_{0 \le k \le N_f} (\mu_k^2-\mu_{N_f+1}^2) 
Z_{RMT}^{(N_f+2)}(m_v+J, m_1,...,m_{N_f},m_v- J).
\ee
If we notice that $\mu_0^2 -\mu_{N_f +1}^2 = 4m_v \Sigma_0 V J + O(J^2)$,
the spectral density is easily obtained from (\ref{spectc})
by differentiating with respect to $J$ and
setting $m_v = i \lambda$. 
After taking the imaginary part we obtain for the microscopic limit
 of the spectral density defined in (\ref{rhosu}),
\be
\rho_S(u) =  \frac 12 |u| 
\prod_{1 \le k \le N_f} (\mu^2_k+u^2) 
\frac{Z_{RMT}^{(N_f+2)}(iu,\mu_1,\cdots,\mu_{N_f},iu)}
{Z_{RMT}^{N_f}(\mu_1,\cdots,\mu_{N_f})},
\label{damg}
\ee
where $u = \lambda V \Sigma_0$. The denominator
arises from the normalization $Z^0_{pq}(J=0)$ of the 
partially quenched partition function. 
The overall normalization is 
such that $\rho(\lambda) \rightarrow \Sigma_0/\pi$ for $ \lambda \rightarrow
\infty$.
This result coincides with the expression first 
derived by Damgaard \cite{Dampart}.
Our derivation provides a natural explanation of their result. The two 
extra flavors are a valence quark and its supersymmetric partner. 
Remarkably, in the context of chRMT, the spectral density
for $\lambda$ not close to zero can still be expressed 
as the ratio of the same two partition functions.
It was shown in \cite{Dampart} that (\ref{damg})
can be reduced to the analytical expression for
the microscopic spectral density of chRMT which,
in the case of zero quark masses, is given by \cite{VZ,V}
\be
\rho_S(u)  = \frac {u}{2} [J^2_{N_f}(u) -J_{N_f+1}(u)
             J_{N_f-1}(u)].
\label{micro2}
\ee

We have thus shown that the microscopic spectral density of chRMT can be
derived from the partition function of partially quenched chiral
perturbation theory. The valence quark mass dependence of the chiral
condensate can be obtained from (\ref{micro2}) by
replacing the sum in (\ref{discval}) by an integral. In terms of the
microscopic variables, $u= \lambda V \Sigma_0$ and $x= m V \Sigma_0$,
this amounts to
\be
\Sigma(x) = \Sigma_0 \int_0^\infty du \frac 1{V \Sigma_0}
\langle \rho(\frac u{V \Sigma_0}) \rangle \frac {2x } {u^2 +x^2}.
\ee
In the thermodynamic limit the spectral density  in this equation can be
replaced by the microscopic spectral density (\ref{micro2}) resulting in
\be
\frac {\Sigma(x)}{\Sigma_0} = x[I_{N_f}(x)K_{N_f}(x)
+I_{N_f+1}(x)K_{N_f-1}(x)].
\label{val}
\ee
By using the duality between flavor and topology \cite{camreview} 
the result for the sector of topological charge $\nu$ is obtained by replacing
$N_f \rightarrow N_f + |\nu|$.

The calculation of
the valence quark mass dependence of the chiral condensate from the
partition function with a noncompact measure
would justify both the choice of the noncompact measure and
the generalization of
the arguments from \cite{Andreev,Seneru} relating the 
compact and the noncompact
measures. 
In a future publication \cite{damgandus} we will come back to this point.
In the next section  we perform a one-loop calculation
of the valence quark mass dependence of the chiral condensate. In section 8
it is shown that  the zero 
momentum component of this result coincides with 
the asymptotic expansion of (\ref{val}) up to this order.

\vskip1.5cm
\noindent
{\bf 7. Valence Quark Mass Dependence of the Chiral Condensate in
Partially Quenched Chiral Perturbation Theory}
\vskip 0.5cm

From a computational point of view, qChPT and pqChPT are not very
different. For example, it is much easier to use the quark
basis~\cite{pqChPT,Sharpe,qChPT,ColPal} with mesons $\bar q_i q_j$ and  
corresponding masses $M^2_{ij}=\Sigma_0 (m_i+m_j)/F^2$ (to lowest order), 
$i,j= s, v$. To lowest order, the bilinear quadratic form in the meson 
fields is diagonal in the off-diagonal 
mesons fields with quark content $\bar q_i q_j$ (with $i \ne j$). 
It is therefore trivial
to write down the propagator for these mesons. 
However, there are 
non-trivial interactions between the diagonal mesons. Nevertheless it
is still possible, as shown by Bernard and Golterman
\cite{pqChPT}, to invert the inverse propagator 
analytically. The result for one valence quark mass is
\cite{pqChPT}: 
\be
G_{ij}(p)=\frac{\delta_{ij}\epsilon_i}{p^2+M^2_{ii}}-
\frac{m_0^2 + \alpha p^2}{(3+N_f \alpha)}
\frac{(p^2+M^2_{ss} )}{(p^2+M^2_{ii})(p^2+M^2_{jj})(p^2+M_{\eta'}^2 )},
\label{eq:neutr}
\ee 
where $\epsilon_i$ is $1$ for the usual mesons and $-1$ otherwise.
The mass of the super-$\eta'$ is given by: 
\be
M^2_{\eta'}=\frac{M^2_{ss}+N_f m_0^2/3}{1+N_f \alpha/3}=M^2_{ss} \frac{1+N_f
y}{1+N_f \alpha/3}, 
\ee 
where $y=m_0^2/3 M^2_{ss}$~\cite{GolLeung}.

The quenched limit is obtained in the limit of infinite sea quark masses
at fixed values of the valence quark mass and $m_0$. On the other hand,
standard ChPT can be recovered in two ways. First, in the sector of the
mesons with two sea quarks, 
in the limit where the sea quark masses are much less than
the mass of the super-$\eta'$ particle. In that limit
the super-$\eta'$ can be integrated out of the effective partition
and standard chiral perturbation theory with $N_f$ flavors is recovered.
The propagator of the mesons
corresponding to the sea quark masses is given by
\be
G_{ss}(p)=\frac{1}{p^2+M^2_{ss}}-\frac{1}{N_f}
\frac{1}{p^2+M^2_{ss}},
\ee
which is just the standard propagator in a quark basis with the singlet
channel projected out. Second, we consider the limit in which the
valence quark masses and the sea quark masses are equal and both are 
much less than  the super-$\eta'$ mass.  
Then, the super-$\eta'$ mass decouples from the theory and, 
to one-loop order, the 
contribution from mesons containing valence quarks 
is cancelled by the superpartners. We again
recover chiral perturbation theory for $N_f$ flavors in both the valence quark
sector and the sea quark sector.

In this work, we are mainly interested in the nonanalytic terms in 
the quark masses
contributing to the chiral condensate at one-loop level.
We will ignore the corrections coming from
the $O(p^4)$ effective Lagrangian and its unknown effective coupling
constants. We will consider the limit where the valence quark mass and
the sea quark masses are much less than the $\eta'$ mass. Therefore, 
contributions from $\Phi_0$
loops proportional to $\log M_{\eta'}^2$ will be ignored as well
\cite{GolLeung}.

The thermodynamic limit of the valence quark mass dependence
was already calculated by Golterman and Leung \cite{GolLeung}. 
Here, we consider the generalization of their result
to  finite volume. The volume dependence of
the chiral condensate in standard chiral perturbation theory
has already received a great deal of attention in the 
literature~\cite{GaL,HasLeu}. 
We refer to these works for  technical details. 
In order to obtain the volume dependence of the quark condensate to one-loop, 
essentially one has to replace the infinite volume pion propagator by 
\be 
\Delta(M^2)= \frac  1V \sum_p   \;
\frac{1}{p^2+M^2},
\ee
where the sum is over the momenta in the box. In the thermodynamic limit
the sum can be replaced by an integral resulting in (in four Euclidean
dimensions)
\be
\Delta(M^2) =\frac{1 }{16 \pi^2} M^2 {\rm ln}
\frac{M^2}{\Lambda^2}, 
\ee 
where $\Lambda$ is the cutoff.

The generalization of the one-loop result 
for the valence quark mass dependence of the chiral
condensate (denoted by a subscript $v$) to finite volume is then given by
\be 
\Sigma_v&=&\Sigma_0 \Bigg[ 1-\frac{1}{F^2} \Bigg\{ N_f
\Delta(M_{sv}^2)-\frac{\alpha+3 y^2 N_f}{3 (1+N_f y)^2} \Delta(M_{vv}^2)
\nonumber \\ &&\hspace{2.2cm} + 
\Big(M^2_{ss} \frac{y}{1+N_f y}-M^2_{vv}
\frac{\alpha+3 N_f y^2}{3 (1+N_f y)^2} \Big) \; \partial_{M_{vv}^2}
\Delta(M_{vv}^2) \Bigg\} \Bigg].\nonumber \\
\label{eq:qq} 
\ee
Notice that $
\partial_{M^2} \Delta(M^2)=-\frac 1V \sum_p
\; {1}/{(p^2+M^2)^2}$. We study the valence quark mass dependence of $\Sigma_v$
for several special cases.
The first  limit of interest is the case that both the valence quark mass and
the sea quark masses are much less than the $\eta'$ mass. This corresponds
to taking the limit $y \rightarrow \infty$ of the expression obtained
in (\ref{eq:qq})
\be 
\Sigma_v = \Sigma_0 \Bigg[
1-\frac{1}{N_f F^2} \Bigg\{ N^2_f \Delta(M_{sv}^2)- \Delta(M_{vv}^2) 
 + (M^2_{ss}-M^2_{vv}) \; \partial_{M_{vv}^2}
\Delta(M_{vv}^2) \Bigg\} \Bigg] .
\label{generalsigma}
\ee
In the thermodynamic limit this reduces to
\be
\Sigma_v\simeq \Sigma_0 \Bigg[
1-\frac{\Sigma_0}{16 \pi^2 F^4 N_f} \Bigg\{ N^2_f (m_v+m_s) \log
\frac{m_v+m_s}{2 \mu} + 2 (m_s-2 m_v) \log \frac{m_v}{\mu} \Bigg\}\Bigg ],
\label{eq:sigQCD} 
\ee 
where $\mu = \Lambda^2 F^2/2\Sigma_0$.
Notice that the valence quark mass occurs in the operator. 
Three special cases of this result are to be mentioned. For $m_s = m_v$
we recover the one-loop result for standard chiral perturbation theory. 

For valence quark masses in the range $m_s \ll m_v \ll m_0$ the valence
quark mass dependence of $\Sigma_v$ is given by
\be
\Sigma_v \simeq \Sigma_0 \left [
1-\frac{\Sigma_0}{16 \pi^2 F^4 } \frac {N_f^2 -4}{N_f} m_v \log
\frac{m_v}{\mu}\right ]. 
\label{press} 
\ee 
Below we will show that this result leads to the expression  
for the slope of the spectral density derived in
\cite{Smilga-Stern}.

For valence quark masses in the range $m_v \ll m_s \ll m_0$
the expression (\ref{eq:sigQCD}) reduces to
\be
\Sigma_v\simeq \Sigma_0 \Bigg[
1-\frac{\Sigma_0}{16 \pi^2 F^4 N_f} \Bigg\{ 
2m_s \log \frac{m_v}{\mu} \Bigg\}\Bigg ].
\label{casequenched} 
\ee 
The expectation is that on this scale for the valence quark mass 
the result for $\Sigma_v$
should reduce to the quenched result. 

Let us consider the quenched limit of (\ref{eq:qq}). 
This limit amounts to 
taking the sea quark mesons much heavier than the $\eta'$ mass, i.e. the
limit $y\rightarrow 0$ while
$M_{ss}^2 y = m_0^2/3$. In this limit (\ref{eq:qq}) reduces to
\be 
\Sigma_v^Q&=&\Sigma_0
\Bigg[ 1-\frac{1}{3 F^2} \Bigg\{ -\alpha \Delta(M_{vv}^2)+ \Big(
m_0^2-\alpha M_{vv}^2  \Big) \; \partial_{M_{vv}^2} \Delta(M_{vv}^2)
\Bigg\} \Bigg] 
\label{eq:Qqc} 
\\ &\simeq& \Sigma_0 \Bigg[
1-\frac{1}{48  \pi^2 F^2}  \Big( m_0^2-2 \alpha M_{vv}^2  \Big) \log
\frac{M^2_{vv}}{\Lambda^2} \Bigg] \quad {\rm for } \quad V\rightarrow \infty, 
\ee
where, in the second line, we have only quoted the infrared singular terms.
For a valence quark masses much less than $m_0$, the term proportional to 
$\alpha$ can be ignored. 

Let us return to (\ref{casequenched}). We observe that it corresponds to
the quenched result  provided that  $m_0^2 = 3 M_{ss}^2/ N_f$. 
Indeed, for the low-momentum
modes we expect that, according to the Witten-Veneziano formula, the
topological contribution to the $\eta'$ mass is determined by the global
topological susceptibility. As dictated by the chiral Ward identity,
this quantity is suppressed by  the screening of topological charge due
to  light sea quarks and is given by \cite{Shifman}
\be
 \langle \nu^2 \rangle &=&  \frac {F^2 M_{ss}^2}{2N_f}.
\label{screen}
\ee
If we combine this result with the relation between $m_0$ and 
$\langle \nu^2\rangle$
given in (\ref{WV}) we recover the previously mentioned condition.

\vskip1.5cm
\noindent
{\bf 8. From pqChPT to chRMT}
\vskip 0.5cm 

Let us study the valence quark mass dependence of $\Sigma_v$
in the chiral limit of the sea quark masses and valence quark masses 
in the domain 
\be
\lambda_{\rm min}\ll m_v \ll \frac 1{L^2\Lambda_{\rm QCD}}.
\label{overlapdomain}
\ee
As has been argued in section 3, in this domain both
 the random matrix expressions (which follow from
the zero momentum component of the partition function (\ref{superpart})) 
and the pqChPT results
(\ref{eq:qq}) are valid. For $m_v$ in the range (\ref{overlapdomain})
the propagator $\Delta(M^2_{vv})$ is dominated 
by the zero momentum mode resulting in
\be
\Delta(M^2_{vv}) \sim \frac 1{M_{vv}^2 V}.
\ee
The   valence quark mass dependence then reduces to 
\be
\Sigma_v &\sim& \Sigma_0(1 - \frac {N_f} {V M_{sv}^2 F^2})\nonumber \\
         &\sim &  \Sigma_0(1 - \frac {N_f} { m_vV \Sigma_0}).
\label{intermediate}
\ee

In section 6 we derived the valence quark mass dependence of the chiral 
condensate from the 
zero momentum pqChPT partition function. The result that was obtained
by integrating the microscopic spectral density coincides with the expression
found from chRMT \cite{vPLB}. The asymptotic expansion 
in powers of  $1/x\equiv 1/m_vV\Sigma_0$ of $\Sigma_{RMT}(x)$
given in (\ref{val}) with the replacement $N_f \rightarrow N_f +|\nu|$
 is given by 
\be
\frac {\Sigma_{RMT}(x)}{\Sigma_0} \sim 1- \frac {N_f+|\nu|}{x} +
\frac{(N_f+|\nu|+\frac 12)(N_f+|\nu|-\frac 12)}{2x^2} + O(\frac 1{x^3}).
\label{asymptote}
\ee
 In the chiral limit the global topological susceptibility  is
completely screened. For $\nu =0$ we thus find
\be
\Sigma(x) \sim \Sigma_0 \left (1 - \frac {N_f}{m_v V \Sigma_0}\right ),
\ee
which coincides with the result for chiral perturbation theory
in this domain.

In the intermediate region where $M_{vv} L \sim 1$, pqChPT is still 
meaningful as a perturbation theory. The finite volume effects 
are embedded in the pion 
propagator. For the propagator of a free scalar particle in a finite
volume we have \cite{GL,HasLeu}, 
\be
\Delta(M^2)&=&\frac{1}{16 \pi^2} M^2 \log \frac{M^2}{\Lambda^2}+ g_1(M^2,L),
\nonumber\\
\partial_{M^2} \Delta(M^2)&=&\frac{1}{16 \pi^2}(1+ \log\frac{M^2}{\Lambda^2})
- g_2(M^2,L),
\label{deltag}
\ee 
where
\be
g_r(M^2,L) = \frac 1{16\pi^2} \int_0^\infty d\lambda \lambda^{r-3}
\sum_{{\bf n}\ne 0} e^{-\lambda M^2 - {\bf n}^2 L^2/4\lambda},
\ee
and the sum is over a four dimensional lattice of integers.
For $ 1/\Lambda_{QCD} L^2 \ll M \ll  1/L$ the functions $g_r$ can be
expanded in powers of $M$ resulting in
\be
g_1(M^2,L) &=& \frac{1}{M^2 L^4}-\frac{\beta_1}{L^2}+\frac {\beta_2}2 M^2 
+O(M^4), \nonumber\\
g_2(M^2,L) &=&\frac{1}{M^4 L^4}+\frac{1}{16 \pi^2} \log (M^2 L^2)-
\frac {\beta_2}2+ O(M^2),
\label{gexpansion}
\ee
where $\beta_1=0.140461$ and $\beta_2=-0.020305$ are shape coefficients of a 
$4$-dimensional cubic box \cite{HasLeu}. 
Let  us consider the chiral limit $m_s \rightarrow 0$. 
Then, in the intermediate range, the asymptotic result 
(\ref{intermediate}) is reproduced.
For $ML \sim 1$ the expansion (\ref{gexpansion}) breaks down and 
the functions $g_r$ have to be evaluated 
numerically.

\begin{center}
\begin{figure}[!ht]
\centering\includegraphics[width=75mm]{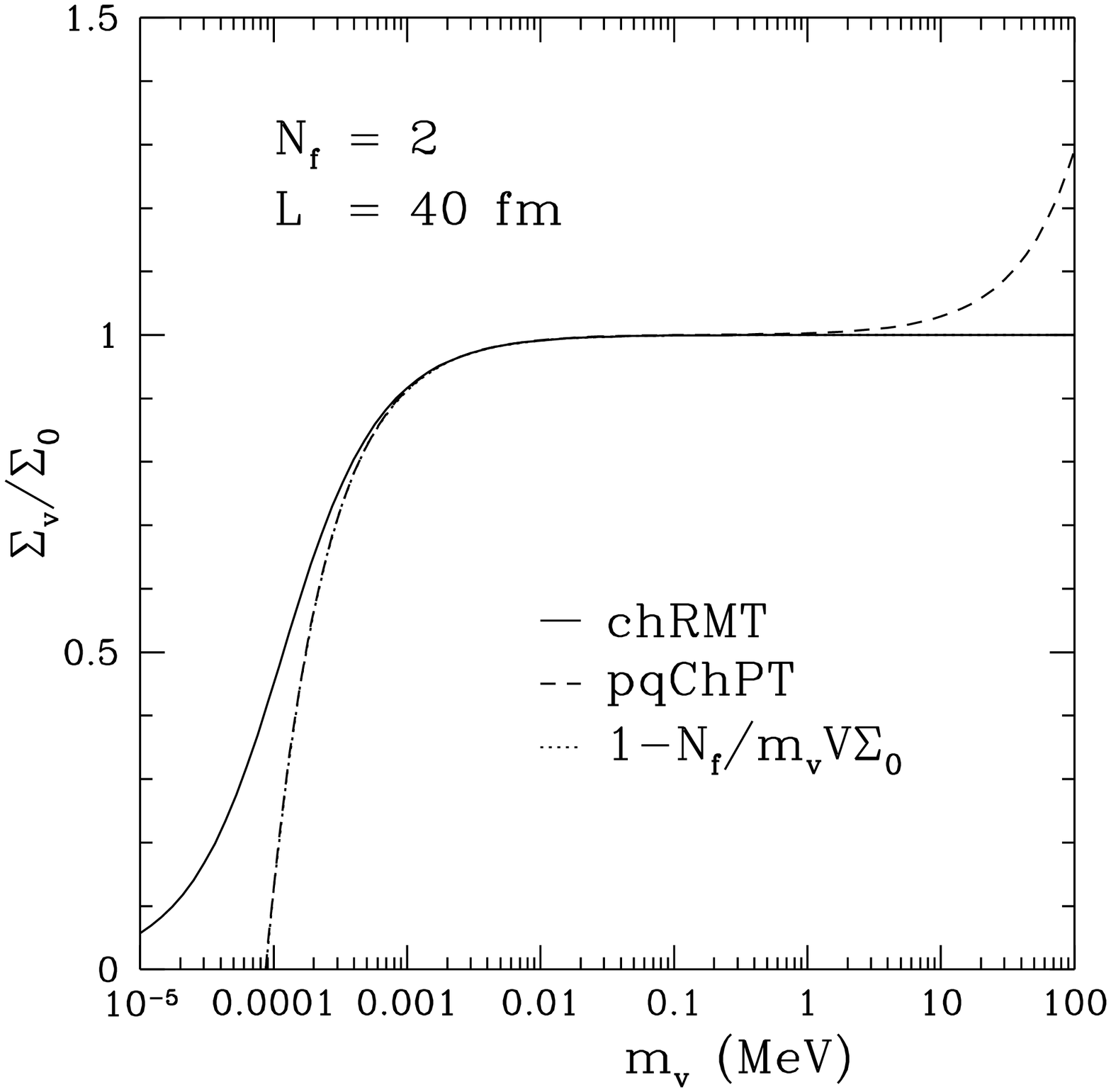}
\centering\includegraphics[width=75mm]{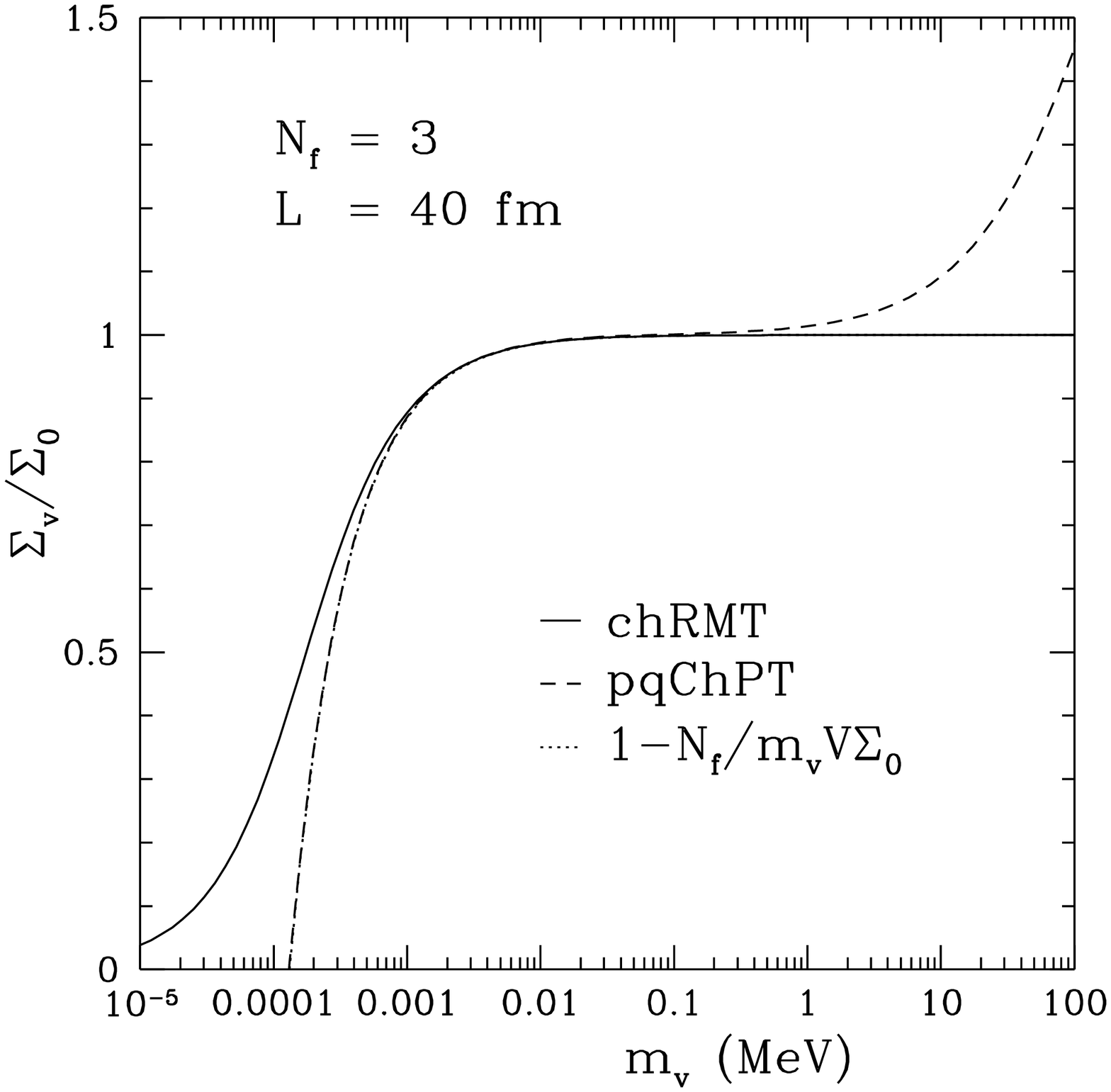}
\vspace*{0.3cm}
\begin{center}
\begin{minipage}{13cm}
\baselineskip=12pt
{\begin{small}
Fig. 1. The valence quark mass dependence of the chiral condensate
$\Sigma_v/\Sigma_0$ as a function of the 
valence quark mass in $MeV$ for two (left) and three (right) massless flavors. 
\end{small}}
\end{minipage}
\end{center}
\vspace*{-0.1cm}
\end{figure}
\end{center}

In Fig. 1 we show the ratio $\Sigma_v/\Sigma_0$ as a function of the
valence quark mass in $MeV$ for two (left) and three (right) massless 
flavors for a box of size $(40\, fm)^4$. 
The dashed curves represent the general
expression for the valence quark mass dependence given in (\ref{generalsigma})
in the limit $m_s \rightarrow 0$ and $\Delta(M^2)$ and 
$\del_{M^2}\Delta(M^2)$ given by (\ref{deltag}) with
the exact numerical result for the functions $g_r$. If the numerical results
for the $g_r$ are replaced by the leading term of 
the small mass expansion (\ref{gexpansion})
we obtain the dotted curves (which first lie almost exactly underneath 
the pqChPT and then underneath the chRMT curves). 
The random matrix result for the valence quark
masses (\ref{val}) 
for $N_f = 2 $ or $N_f = 3$ and $\nu = 0$ is given by the solid curves.
We stress that this result can also be derived from the 
pqChPT partition function as shown in section 6.
The results in this figure have been calculated for 
$\Lambda = 770 \, MeV$, $F= 93\, MeV$ and 
$\Sigma_0/F^2 = 1660\, MeV$. For these values the Thouless energy
is $E_c \equiv F^2/\Sigma_0 L^2 \approx 0.015 \, MeV$.
We observe a region
where the random matrix results and the pqChPT results are in agreement. The
size of this region is consistent with the range (\ref{valdomain}).
For $N_f = 2$ this region appears larger than for $N_f = 3$ because the
leading correction vanishes in the thermodynamic limit for $N_f = 2$.
Because of the large box size, the $1/L^2$ terms 
are almost invisible in this figure at the point where the
nonzero momentum modes become important. 

\begin{center}
\begin{figure}[!ht]
\centering\includegraphics[width=75mm]{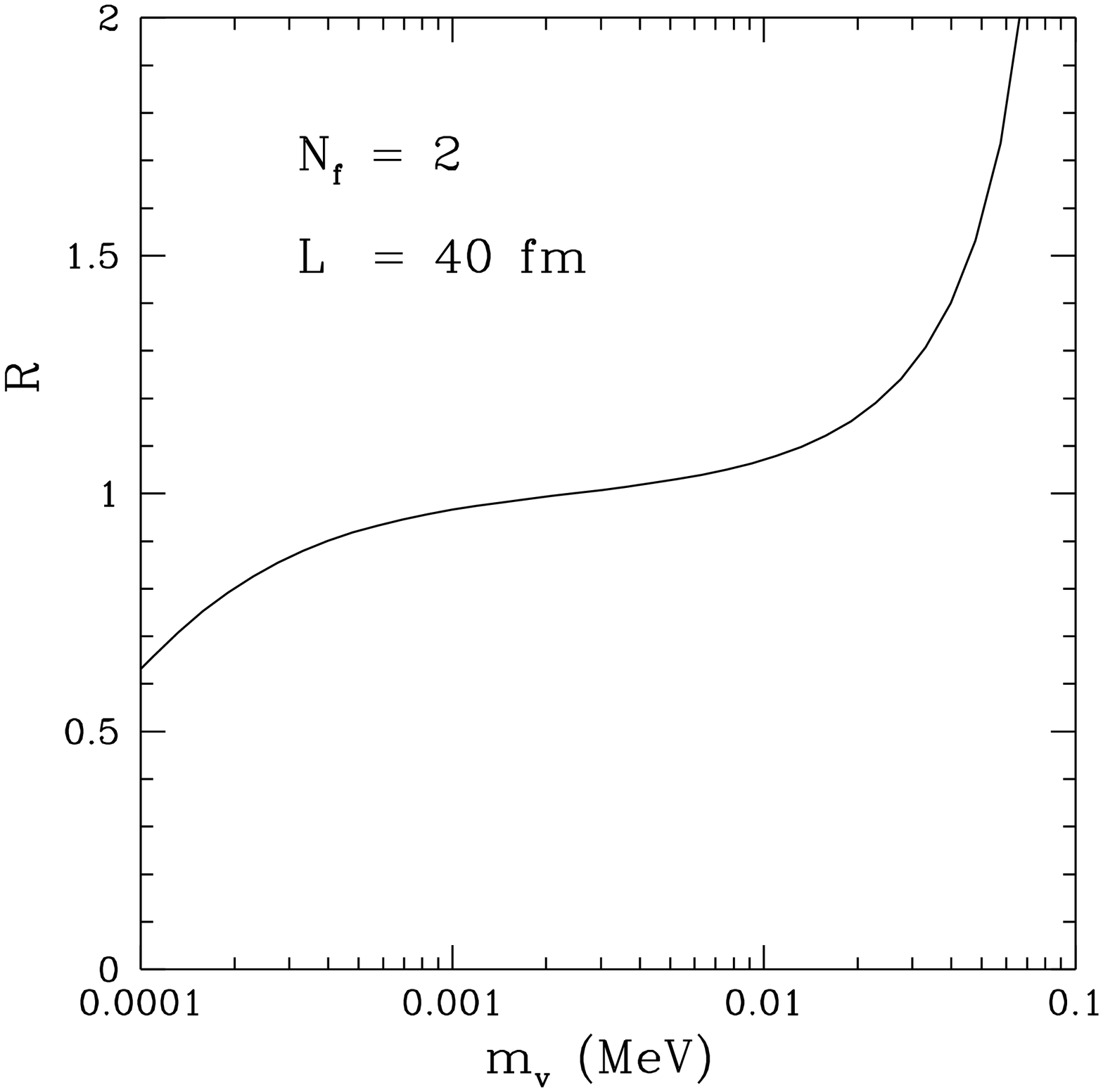}
\centering\includegraphics[width=75mm]{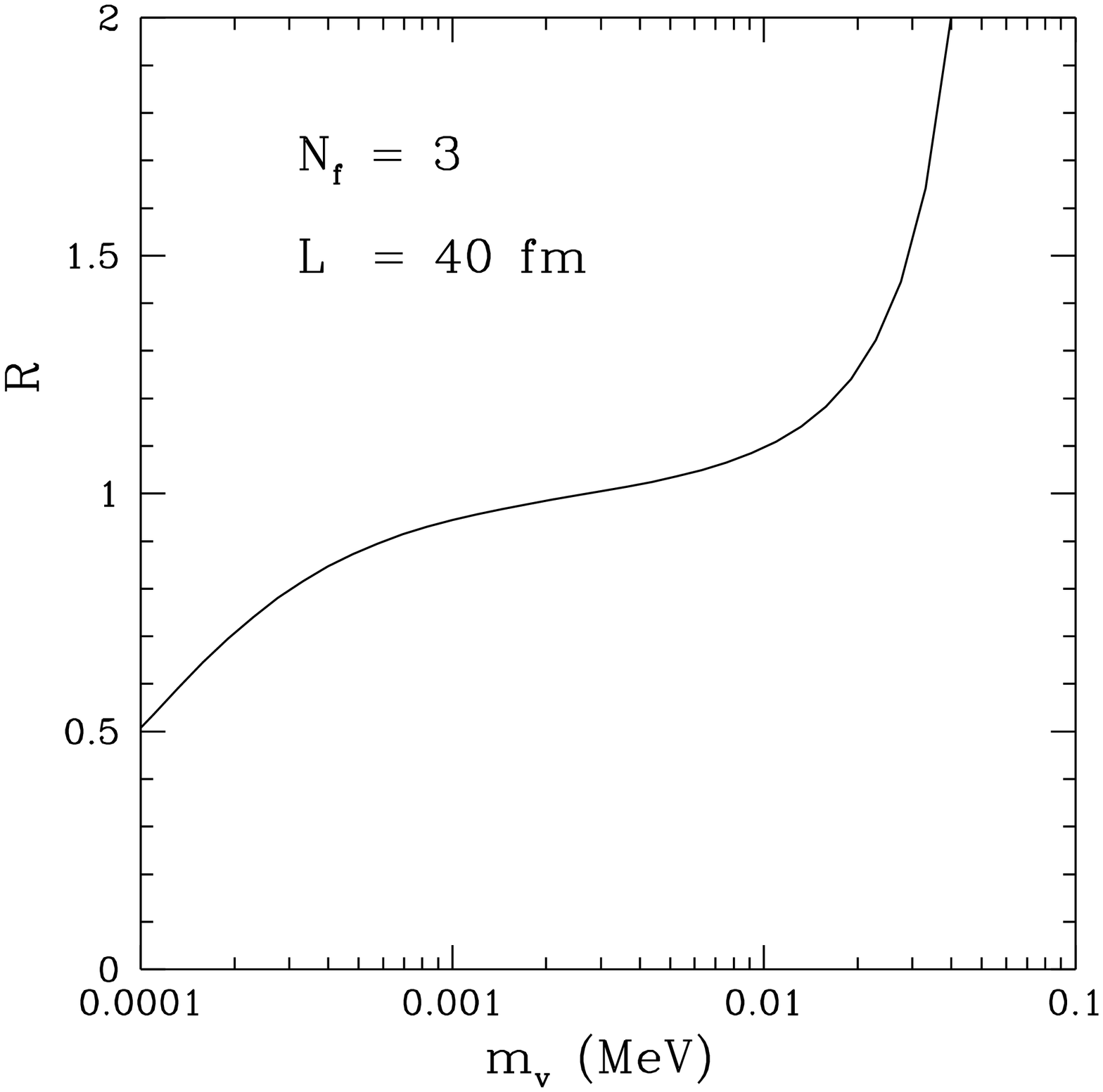}
\vspace*{0.3cm}
\begin{center}
\begin{minipage}{13cm}
\baselineskip=12pt
{\begin{small}
Fig. 2. The  
ratio $R = ({\Sigma_{RMT} -\Sigma_0})/({\Sigma_v -\Sigma_0})$
versus the valence quark mass $m_v$
for two and three massless flavors.
\end{small}}
\end{minipage}
\end{center}
\vspace*{-0.1cm}
\end{figure}
\end{center}                     

In order to study the point where the nonzero momentum modes become important
we study the ratio
\be
R = \frac{\Sigma_{RMT} -\Sigma_0}{\Sigma_v -\Sigma_0}.
\ee
According to (\ref{intermediate}) and (\ref{asymptote}) this ratio is   
close to 1
in the range (\ref{overlapdomain}). In Fig. 2 we plot this ratio as
a function of the valence quark mass $m_v$ for a box size of $(40\, fm)^4$
and $N_f =2$ (left) or $N_f = 3$ (right). At intermediate masses we
indeed observe a plateau where chRMT and pqChPT are in agreement.
Notice that 
in the center of the overlap region (\ref{overlapdomain}), where
$m_v$ scales as $1/ L^3$, 
the deviation of this ratio from 1 is of order $1/L$. 
Indeed, for larger volumes we have observed a
more pronounced plateau. The
Thouless energies can be read off from these pictures.
For $N_f = 2$ and $N_f = 3$, we find values of around  0.01-0.02 $MeV$
which is in good agreement with the
estimate from $E_c = F^2/\Sigma_0 L^2$.

\vskip1.5cm
\noindent
{\bf 9. Spectral Density in the Diffusive Domain}
\vskip 0.5cm

As already has been remarked in section 1, 
the spectral density is obtained from the
discontinuity of the valence quark mass dependence of the chiral condensate
along the imaginary axis.
First, notice that the quark mass always occurs in the renormalization group
invariant combination
$m \langle \bar q q \rangle$ with the sign of $\langle \bar q q \rangle$
opposite to the sign of $m$. This combination can be rewritten as
$|m| \Sigma$. For similar reasons, $\Sigma_0$ in the expressions for
the valence quark mass dependence of the chiral condensate is actually
multiplied by $|m_v|/m_v$ with $|m_v| = \sqrt{ m_v^2}$. From the elementary
identity
$\lim_{\epsilon \rightarrow 0}
[\sqrt(-x^2 + i\epsilon)-\sqrt(-x^2 - i\epsilon)]/ix = 2$ we thus conclude that
the discontinuity of the first term
in (\ref{eq:sigQCD}) is equal to 2. According to (\ref{spectdisc})
this  results in a contribution to the spectral
density of $\Sigma_0/\pi$ which is in agreement with the Banks-Casher  
formula (\ref{Banks-Casher}). 
The discontinuities that enter
in the calculation of the spectral density from the one-loop result for the
valence quark mass dependence of $\Sigma_v$ are given by 
\be
&&\left .{\rm Disc}\right |_{m_v = ix} 
\left ( \frac {|m_v|}{m_v}\right)  = 2, \\
&&\left .{\rm Disc}\right |_{m_v = ix}\left 
( \frac {|m_v|}{m_v} \log(m_s + |m_v|) \right )  = \log( m^2_s + x^2),\\
&&\left .{\rm Disc}\right |_{m_v = ix}\left (  \log(m_s + |m_v|) \right)  =
2i \arctan \frac x{m_s},\\
&&\left .{\rm Disc}\right |_{m_v = ix}  \log|m_v|  = i \pi.
\ee
The spectral density can be calculated
from the one-loop
result for the valence quark mass dependence of $\Sigma_v$ (\ref{eq:sigQCD}) 
 by means of 
$\langle\rho(\lambda)\rangle/V = \left .{\rm Disc}\right |_{m_v = 
i\lambda}\Sigma_v(m_v)/2\pi$. The result is
\be 
\frac{\langle\rho (\lambda)\rangle_{m_s}}V
&=&\frac{\Sigma_0}{\pi} \Bigg[ 1+\frac{\Sigma_0}{32 \pi^2 N_f F^4 }
\Bigg\{ 2 N^2_f |\lambda| {\rm Arctg} \frac{|\lambda|}{m_s}-4 \pi
|\lambda| -N^2_f m_s \log
        \frac{\lambda^2+m^2_s}{\mu^2}-4 m_s \log \frac{|\lambda|}{\mu} \Bigg\}
\Bigg] .\nonumber\\
\label{rhomlam}
\ee
For $m_s \ne 0$ the spectral density diverges logarithmically.
In the limit $\lambda \gg m_s$ this expression reduces to 
\be 
\frac{\langle \rho(\lambda)\rangle_{m_s}}V &=&\frac{\Sigma_0}{\pi}
\Bigg[ 1+\frac{ (N^2_f-4) \Sigma_0}{32 \pi^2 N_f F^4 }  |\lambda|\Bigg],
\label{eq:rhoQCD} 
\ee
which was first obtained by Smilga and Stern \cite{Smilga-Stern}.

Our expression for  the spectral density can be used to evaluate the connected
contribution to the integrated three-point scalar correlator,
\be
K^{abc} = \int d^4 x d^4 y d^4 z \langle 0|S^a(x) S^b(y) S^c(z)|0\rangle,
\ee
with $S^a = \bar q t^a q$ and $t^a$ is one of the generators of the flavor
$SU(N_f)$ group in the normalization ${\rm Tr} (t^a t^b) = \delta^{ab}/2$.
This correlator can be expressed in terms of the spectral density by
\cite{Smilga-Stern}
\be 
K^{abc} = -m_s d^{abc} \int_0^{\infty} d \lambda \;
\frac{m^2_s-3 \lambda^2}{(m^2_s+\lambda^2)^3}\langle \rho(\lambda) 
\rangle_{m_s},
\ee
where $d^{abc}$ are the symmetric structure constants of $SU(N_f)$.
By substitution of the expression (\ref{rhomlam}) 
for the spectral density  we then obtain
\be
\frac 1V K^{abc} = d^{abc}\frac{N^2_f-12}{N_f m_s}
\frac {\Sigma^2_0}{128\pi^2 F^4},
\ee
which is in complete agreement with the result obtained by Smilga and Stern
using standard ChPT \cite{Smilga-Stern}. Notice that both 
terms of order $O(m^0_s)$ and of $O(m^1_s)$ in (\ref{rhomlam}) contribute to
this result. The existence of contributions of this type 
was already pointed out in 
\cite{Smilga-Stern}, but partially 
quenched chiral perturbation theory enabled us
to actually calculate them.

The spectral density in the quenched case can be obtained in exactly the same
way by calculating the discontinuities of the valence quark mass dependence
across the imaginary axis.  The result is
\be
\frac{\langle \rho^Q(\lambda) \rangle}V = \frac {\Sigma_0}\pi\left [ 1 - 
\frac {m_0^2}{48\pi^2 F^2} \log \frac {|\lambda|} \mu - 
\frac{\Sigma_0\alpha}{24 \pi F^4} |\lambda|\right ].
\ee
A logarithmic divergence of the spectral density in the quenched limit has
indeed been observed in quenched instanton liquid simulations \cite{Osborn}.
This result allows us to interpret the coefficient of $\log |\lambda|$
in the limit $\lambda \ll m_s$ of (\ref{rhomlam})
in terms of the screened topological 
charge (\ref{screen}).

\vskip 1.5cm
\noindent
{\bf 10. Conclusions}
\vskip 0.5cm

We have studied the spectral density of the QCD Dirac operator by
means of the valence quark mass dependence  of the chiral condensate
in partially quenched Chiral Perturbation Theory (pqChPT). This is 
the extension of standard ChPT with an additional valence quark. Its mass
is required to probe the Dirac spectrum.
We have considered 
two different domains of applicability: the ergodic domain 
with $m_v \ll F^2/L^2\Sigma_0$ and the diffusive domain with
$F^2/L^2\Sigma_0\ll m_v \ll \Lambda_{QCD}$. The energy scale 
$F^2/L^2\Sigma_0$ is the equivalent of the Thouless energy in mesoscopic
physics.
In the ergodic domain the partition function is dominated by the zero momentum
modes and can be reduced to chiral Random Matrix Theory (chRMT).
In particular, we have shown this for the microscopic spectral density, the
valence quark mass dependence of the chiral condensate and the
asymptotic tail of the two-point spectral correlation function. 
We have found that
a noncompact superunitary measure is required to directly 
reproduce the correct
valence quark mass dependence. These results strongly 
suggest that the zero momentum
component of the pqChPT partition function is equivalent to the chRMT
partition function. 

The diffusive regime is the realm of Chiral Perturbation Theory.
We have reproduced the Smilga-Stern expression for the slope of the
spectral density  of  the QCD Dirac operator and have calculated its first
order  correction in the quark mass. We have shown that these results 
are consistent with standard ChPT. For nonzero sea quark masses the 
spectral density diverges logarithmically with a coefficient that is
proportional to the screened topological susceptibility.

We have studied the transition from chRMT to pqChPT in the domain
where both are applicable and have found that both theories give the
same valence quark mass dependence of the chiral condensate in this domain.
The Thouless energy can thus be derived from the finite volume
scalar propagator.

If pqChPT provides an accurate description of the QCD partition function
we have shown that up to the Thouless energy 
the Dirac spectrum can be described
by chRMT. Agreement with lattice QCD simulations suggests that this is
indeed the case. 

Other scenarios are possible. For example,
the proof of the Vafa-Witten theorem \cite{Vafa} 
is not valid for imaginary quark masses.
Therefore, breaking of "isospin" symmetry between the 
sea quark sector and the valence 
quark sector cannot be excluded. Related to this is the possibility that
the pion decay constant is different for the sea quark channels and the
valence quark channels.  This might restrict the domain of validity of chRMT.
However, universality arguments suggest that such scenarios are unlikely. 
Ultimately, these and other dynamical 
questions have to be answered by means of lattice QCD simulations. 

\vskip 1.5cm
\noindent
{\bf Acknowledgements}
\vskip 0.5cm

This work was partially supported by the US DOE grant
DE-FG-88ER40388. One of us (D.T.) was supported by
Schweizerischer Nationalfonds.
T. Guhr, A. Sch\"afer, A. Smilga, M. Stephanov, H.A. Weidenm\"uller 
and T. Wettig are acknowledged for useful discussions. We particularly
benefitted from many useful comments by P. Damgaard.

\vskip 1.5cm
\noindent
{\bf Appendix A. Calculation of the Microscopic Spectral Density for $N_f =1$}
\vskip 0.5cm

In this appendix we calculate the integrals in the calculation
of the microscopic spectral density from (\ref{zcgen}). We explicitly
work out the integrals for $N_f =1$ and indicate how they can be performed
for arbitrary $N_f$.
For $N_f = 1$ the different factors contributing to the integral are
given by
\be
B(\Lambda) &=& \frac{(e^{i\theta_0} - e^{i\theta_1})}{(e^{i\theta_0} - 
e^{i\theta_2})(e^{i\theta_1} - e^{i\theta_2)}},\\
B(\cos \theta) &=& \frac{(\cos\theta_0 - \cos \theta_1)}
{(\cos\theta_0 - \cos \theta_2)(\cos\theta_1 - \cos \theta_2)},\\
B(M_k)&=& \frac{(m_v-m_s)}{2J\Sigma_0 V (m_v-m_s)} = \frac 1{2J\Sigma_0 V}.
\ee
In the chiral limit $m_s \rightarrow 0$, the determinant can be written as
\be
\det e^{M_k V \Sigma_0\cos\theta_l} = e^{m_v V\Sigma_0 \cos\theta_0} -
e^{m_v V\Sigma_0 \cos\theta_1}.
\ee
Combining all factors we find
\be
\left . \frac 1V  \del_J Z^C(J)\right |_{J=0} = \frac{\Sigma_0}{8\pi^3} 
\int d\theta_0 d\theta_1 
d\theta_2 
\frac{|e^{i\theta_0} - e^{i\theta_1}|^2 
(\cos\theta_0 - \cos \theta_2)(\cos\theta_1 - \cos \theta_2)}
{|e^{i\theta_0} - e^{i\theta_2}|^2|e^{i\theta_1} - e^{i\theta_2}|^2 
(\cos\theta_0 - \cos \theta_1)}e^{m_v V\Sigma_0 (\cos\theta_0-\cos\theta_2)},
\nonumber \\
\ee
where we have used the anti-symmetry of the measure in $\theta_0$ and 
$\theta_1$. 

We first perform the integral over $\theta_1$
\be
I_1 \equiv \int d\theta_1 \frac{|e^{i\theta_0} - e^{i\theta_1}|^2 
(\cos\theta_1 - \cos \theta_2)}
{(\cos\theta_0 - \cos \theta_1)|e^{i\theta_1} - e^{i\theta_2}|^2}.
\ee
The integrand can be simplified with the help of the identity (\ref{identity}) 
(for  $s = i\phi$) resulting in
\be
I_1 = \int d\theta_1 \frac{(e^{-i\theta_0} - e^{-i\theta_1}) 
(1- e^{-i\theta_1}e^{-i\theta_2})}
{(1- e^{-i\theta_0}e^{-i\theta_1})(e^{-i\theta_1} - e^{-i\theta_2})}.
\ee
To evaluate the integral we add opposite  infinitesimal imaginary
increments to $\theta_0$ and $\theta_2$ and take the 
limit after the integration. For definiteness, let us add a negative
imaginary increment to $\theta_0$. 
Then we can expand the appropriate
terms in a geometric series. We observe that $\theta_1$ only occurs as
positive powers of $\exp(-i\theta_1)$, and only the constant
term contributes. The result for the integral is thus
\be
I_1 = -2\pi e^{i\theta_2 -i \theta_0}.
\ee
For the opposite choice of the imaginary
increment we obtain the complex conjugate expression resulting in the
same final answer,
\be
\left . \frac 1V  \del_J Z(J)\right |_{J=0} = -\frac{\Sigma_0}{4\pi^2} 
\int d\theta_0 d\theta_2 e^{i\theta_2-i\theta_0}
\frac{(\cos\theta_0 - \cos \theta_2)}
{|e^{i\theta_0} - e^{i\theta_2}|^2}e^{x(\cos\theta_0-\cos\theta_2)},
\ee
where $x = m_v V\Sigma_0$. 
Notice that only the real part of $I_1$ contributes to the integral over
$\theta_2$ and $\theta_0$. 
It is simplest to evaluate this integral by
first differentiating it with respect to $x$ and using once more
 the identity (\ref{identity})
\be
\del_x \Sigma(x) &=& -\frac{\Sigma_0}{16\pi^2} \int d\theta_2 d\theta_0
(1- e^{-i\theta_0-i\theta_2})(1- e^{+i\theta_0+i\theta_2})
e^{i\theta_2 -i \theta_0}
e^{x(\cos\theta_0-\cos\theta_2)},\nonumber\\
&=& \frac {\Sigma_0} 2 [J_1(-ix)J_1(-ix) +J_0(-ix)J_2(-ix)],\nonumber\\
&=&\frac {\Sigma_0} 2 \del_x[xJ_1^2(-ix) -x J_0(-ix)J_2(-ix)].
\ee
This equation can be integrated trivially with an
integration constant that is determined by the asymptotic limit for the
spectral density. The result for the microscopic spectral density
is obtained by putting $m_v = i\lambda$ and taking the microscopic
limit according to (\ref{rhosu}) (with $u= \lambda V\Sigma_0$),
\be
\rho_S(u) = \frac u2 [J_1^2(u) - J_0(u) J_2(u)],
\ee
in complete agreement with the chRMT result for $N_f = 1$.

For an arbitrary number of flavors $N_f$, we systematically use the 
symmetries of the integrand. First, the determinant $ \det_{0 \le k,l 
\le N_f} [\exp(m_k \cos \theta_l)]$ appearing in (\ref{zcgen}) and the 
Berezinian $B(\cos \theta_i)$ are both antisymmetric in $\theta_i 
\leftrightarrow \theta_j$, for $i \neq j$ and $i,j=0, \cdots , N_f$. 
Therefore the whole expression can be simplified to:
\be
\del_J Z^0_{pq} \sim \int d \theta_0 \; d \theta_{N_f+1} \; \frac{\cos \theta_0
-\cos \theta_{N_f+1} }
{ | e ^{i \theta_0}- e^{i \theta_{N_f+1}}|^2 } e^{x 
(\cos \theta_0-\cos \theta_{N_f+1})} \tau_{N_f} (\theta_{N_f+1},\theta_0),
\ee
where
\be
\tau_{N_f}(\theta_{N_f+1},\theta_0)=\int \prod_{k=1}^{N_f} \; d \theta_k \; 
| \Delta(e^{i \theta_k}) |^2 \; \omega (\theta_{N_f+1},\theta_k) \; 
\omega (-\theta_0,\theta_k).
\ee
and $\omega(\theta,\phi)$ is defined by
\be
\omega(\theta,\phi) =\frac{\cos \theta - \cos \phi}{|e^{i\theta} - 
e^{i\phi}|^2}.
\ee

Considering again $\partial_x \Sigma(x)$ and using the fact that 
$| \Delta(e^{i \theta_k}) |^2$ is the sum of $1$ plus
terms $\exp(i \sum_k n_k \theta_k)$ with $n_k$ finite and
$\sum_k n_k=0$
\cite{LS}, we can show that the only term in $\tau_{N_f} 
(\theta_{N_f+1},\theta_0)$ contributing to the integral is proportional 
to $\exp[i N_f (\theta_0-\theta_{N_f+1})]$. Notice, that 
the term proportional to $\sin[N_f 
(\theta_{N_f+1}-\theta_0)]$ does not contribute to the integral. 
To obtain the final result for the microscopic spectral density
we again use the symmetry properties of the integrand with 
respect to the exchange and 
translations of the $\theta_k$ variables.

\vskip 1.5cm
\noindent
{\bf Appendix B}
\vskip 0.5cm

In this appendix we calculate the Berezinian corresponding to the
diagonalization of a superunitary matrix. The calculation is analogous
to the case of a super-Hermitean matrix \cite{Guhr91,Alfaro}.
We can decompose the superunitary matrix $U$ into its eigenvalues $\Lambda$ and
another superunitary matrix $V$ by $U=V\Lambda V^{\dagger}$.  Since $U$ is
superunitary its eigenvalues are just phases, i.e. $\Lambda={\rm diag}
(e^{i \theta_k})$.
The relevant length element is
\be
|ds|^2 = {\rm Str} dU dU^\dagger =
{\rm Str} \left( d\Lambda d\Lambda^\dagger \right) + {\rm Str} \left( 
\left[V^\dagger dV,\Lambda \right] 
\left[V^\dagger dV,\Lambda \right]^\dagger \right).
\ee
Since we are only interested in the $\Lambda$ dependence of the Berezinian
we can use the parametrization $V=e^{iH}$ so that $V^\dagger dV = i dH$.
We then have
\be
|ds|^2 = \sum_i \epsilon(i) d\Lambda_i d\Lambda_i^* +
\sum_{i,j} \epsilon(i) \epsilon(j) |\Lambda_i-\Lambda_j|^2 
dH_{i,j} dH_{i,j}^*,
\ee
where $\epsilon(i)=1 (-1)$ if the $\Lambda_i$ corresponds to a fermion (boson).
If we write this in the form $|ds|^2 = g_{a,b} dz_adz_b^* $ then the 
Berezinian is given by $\sqrt{{\rm Sdet}(g)}$,
\be
|B(\Lambda)|^2 = \frac{\prod_{i>j} |\Lambda_i^f-\Lambda_j^f|^2
\prod_{i>j} |\Lambda_i^b-\Lambda_j^b|^2}
{\prod_{i,j} |\Lambda_i^f-\Lambda_j^b|^2}.
\ee


\begin{thebibliography}{9}


\bibitem{DeTar}C.~DeTar,
{\it Quark-gluon plasma in numerical simulations of QCD}, in {\it
Quark gluon plasma 2}, R. Hwa ed., World Scientific 1995.
\bibitem{Ukawa}A.~Ukawa, Nucl. Phys. Proc. Suppl. {\bf 53} (1997) 106.

\bibitem{Smilref}A.V. Smilga, Phys. Rep. 291, (1997) 1. 
\bibitem{BC}T.~Banks and A.~Casher, Nucl. Phys. {\bf B169} (1980).
103
\bibitem{SVR}E.V. Shuryak and J.J.M. Verbaarschot,
Nucl. Phys. {\bf A560} (1993) 306.
\bibitem{Vinst}J.J.M. Verbaarschot, Nucl. Phys. {\bf B427} (1994) 434.
\bibitem{Tilo} M.E. Berbenni-Bitsch, S. Meyer, A. Sch\"afer,
J.J.M. Verbaarschot and  T. Wettig, Phys. Rev. Lett. {\bf 80} (1998) 1146.
\bibitem{many}M.E. Berbenni-Bitsch, M. Gockeler, T. Guhr, A.D. Jackson,
J.Z. Ma, S. Meyer, A. Sch\"afer, H.A. Weidenm\"uller, T. Wettig and
T. Wilke,  {\it Crossover to nonuniversal fluctuations in lattice gauge
theory},  hep-ph/9804439.
\bibitem{Guhr-Wilke}T. Guhr, J.Z. Ma, S. Meyer and T. Wilke, {\it Statistical
analysis and the equivalent of a Thouless energy in lattice QCD Dirac spectra},
hep-lat/9806003.
\bibitem{Tilomass}M.E. Berbenni-Bitsch, S. Meyer and T. Wettig, 
hep-lat/9804030. 

\bibitem{Halasz}M.A. Halasz and J.J.M. Verbaarschot,
Phys. Rev. Lett. {\bf 74} (1995) 3920;
M.A. Halasz, T. Kalkreuter and J.J.M. Verbaarschot,
Nucl. Phys. Proc. Suppl. {\bf 53} (1997) 266.
\bibitem{Ma} J.Z. Ma, T. Guhr and T.
Wettig, Eur. Phys. J. {\bf A2} (1998) 87.


\bibitem{markum}R. Pullirsch, K. Rabitsch,
T. Wettig and H. Markum, hep-ph/9803285.
\bibitem{Damgaard}G. Akemann,
P. Damgaard, U. Magnea and S. Nishigaki, Nucl. Phys. {\bf B 487[FS]} (1997)
721. 
\bibitem{brezin}E. Br\'ezin, S. Hikami and A. Zee,
Nucl. Phys. {\bf B464} (1996) 411.
\bibitem{GWu}T. Guhr and T. Wettig, Nucl. Phys. {\bf B506} (1997) 589.
\bibitem{Sener1}A.D. Jackson, M.K. Sener and J.J.M. Verbaarschot, Nucl. Phys.
{\bf B479} (1996) 707.
\bibitem{Seneru}A.D. Jackson, M.K. Sener and J.J.M. Verbaarschot, 
Nucl. Phys. {\bf B506} (1997) 612.
\bibitem{andystudent}K. Splittorff and A.D. Jackson, hep-lat/9805018. 
\bibitem{Tilodam}S.M. Nishigaki, P.H. Damgaard and T. Wettig, hep-th/9803007.
\bibitem{Senerprl}M.K. Sener and J.J.M. Verbaarschot, hep-th/9801042.
\bibitem{microcanonical}D. Callaway and A. Rahman, Phys. Rev. Lett. {\bf 49}
(1982) 613; J. Polony and H.W. Wyld, Phys. Rev. Lett. {\bf 51} (1983) 2257;
J. Kogut, J. Polony, J. Shigemitsu, D.K. Sinclair and H.W. Wyld, Phys. Rev.
Lett. {\bf 53} (1984) 644.
\bibitem{bohigas}O.~Bohigas, M.~Giannoni, Lecture notes in Physics
{\bf 209} (1984) 1; O. Bohigas, M. Giannoni and C. Schmit, Phys. Rev. Lett.
{\bf 52} (1984) 1.
\bibitem{Stern}J. Stern, hep-ph/9801282.
\bibitem{HDgang}T. Guhr, A. M\"uller-Groeling and H.A. Weidenm\"uller,
Phys. Rept. {\bf 299} (1998) 189.

\bibitem{Beenreview}C.W.J. Beenakker, Rev. Mod. Phys. {\bf 69} (1997) 731.
\bibitem{Montambaux}G. Montambaux, in {\it Quantum Fluctuations,
Les Houches, Session LXIII}, 
E. Giacobino, S. Reynaud and J. Zinn-Justin, eds., 
Elsevier Science, 1995, cond-mat/9602071.

\bibitem{Altshuler} B.L. Altshuler, I.Kh. Zharekeshev, S.A. Kotochigova and
B.I. Shklovskii, Zh. Eksp. Teor. Fiz. {\bf 94} (1988) 343.
\bibitem{spreading}H.A. Weidenm\"uller,
Nucl. Phys. A 518 (1990) 1.
\bibitem{Guhr}T. Guhr and A. Mueller-Groeling,  cond-mat/9702113, J. Math.
Phys. (in press).
\bibitem{Imry}N. Argaman, Y. Imry and U. Smilansky,
Phys. Rev. {\bf B47} (1993) 4440.
\bibitem{kravtsovmoriond}V.E. Kravtsov, in
{\it Correlated Fermions and Transport in Mesoscopic Systems},
Les Arcs, 1996, cond/mat9603166.

                                                                 

\bibitem{Altland} A. Altland and Y. Gefen, Phys. Rev. Lett. {\bf 71} (1993)
3339.

\bibitem{Braun} D. Braun and G. Montambaux, Phys. Rev. {\bf 52} (1995) 13903.

\bibitem{Aronov} A.G. Aronov and A.D. Mirlin,
Phys. Rev. {\bf B51} (1995) 6131.
\bibitem{yan}Y.V. Fyodorov and A.D. Mirlin, Phys. Rev. {\bf B51} (1995) 13403.
\bibitem{kravtsov-lerner}V.E. Kravtsov, I.V. Lerner, B.L. Altshuler 
and A.G. Aronov, Phys. Rev. Lett. {\bf 72} (1994) 888.
\bibitem{Altland-Gefen}A. Altland, Y. Gefen and G. Montambaux,
Phys. Rev. lett. {\bf 76} (1996) 1130.

\bibitem{Chalker-kravtsov}J.T. Chalker, V.E. Kravtsov, I.V. Lerner,
JETP Lett. 64 (1996) 386.
\bibitem{shuryak}E. Shuryak, Nucl. Phys. {\bf B302} (1988) 599;
D. Diakonov, 
{\it Talk given at International School of
Physics, 'Enrico Fermi', Course 80: Selected Topics in Nonperturbative QCD}, 
Varenna, 1995, hep-ph/9602375. 

\bibitem{Osbornprl}J.C. Osborn and J.J.M. Verbaarschot, Phys. Rev. Lett. 
(in press) (1998).
\bibitem{Janik} R. Janik, G. Papp, M. Nowak and I.Zahed, hep-ph/9803289.
\bibitem{Osborn}J.C. Osborn and J.J.M. Verbaarschot, Nucl. Phys. {\bf B}
(in press) (1998), hep-ph/9803419.
\bibitem{Morel} A. Morel, J. Physique {\bf 48} (1987) 1111.
\bibitem{pqChPT} C. Bernard and M. Golterman, Phys. Rev. D49 (1994)
486; C. Bernard and M. Golterman, hep-lat/9311070.

\bibitem{GL}J. Gasser and H.~Leutwyler, Phys. Lett. {\bf 188B}(1987) 477;
Nucl. Phys. {\bf B307} (1988) 763.
\bibitem{LS}H.~Leutwyler and A.~Smilga, Phys. Rev. {\bf D46} (1992) 5607. 

\bibitem{V} J. Verbaarschot, Phys. Rev. Lett. {\bf 72} (1994) 2531;
Phys. Lett. {\bf B329} (1994) 351; Nucl. Phys. {\bf B426[FS]} (1994) 
559.

\bibitem{camreview}J.J.M. Verbaarschot, 
{\it Lectures given at NATO Advanced Study Institute on Confinement, 
Duality and Nonperturbative Aspects of  QCD}, Cambridge, 1997, hep-th/9710114.
\bibitem{Christ}S. Chandrasekharan, Nucl. Phys. Proc. Suppl. {\bf 42}
(1995) 475; S. Chandrasekharan and N. Christ, Nucl. Phys.
Proc. Suppl. {\bf 42} (1996) 527; N. Christ, Nucl. Phys. {\bf B} (Proc. Suppl.)
53 (1997) 253.
\bibitem{vPLB}J.J.M. Verbaarschot, Phys. Lett. {\bf B368} (1996) 137.
\bibitem{Trento}J.J.M. Verbaarschot,
in {\it Nonperturbative Approaches to Quantum Chromodynamics},
D. Diakonov, ed., Gatchina 1995.
\bibitem{Andreev}A.V. Andreev,
B.D. Simons, and N. Taniguchi, Nucl. Phys {\bf B432 [FS]} (1994) 487.
\bibitem{Toublan}J.C. Osborn, D. Toublan and J.J.M. Verbaarschot, 
in preparation.
\bibitem{GolLeung}M.F.L. Golterman and K.C. Leung,  hep-lat/9711033;
M.F.L. Golterman, Acta Phys. Polon. {\bf B25} (1994). 

\bibitem{Smilga-Stern}A. Smilga and J. Stern, Phys. Lett. {\bf B318} (1993) 
531.
  
\bibitem{Dampart}P.H. Damgaard, Phys. Lett. {\bf B424} (1998) 322;
G. Akemann and P.H. Damgaard, hep-th/9802174;
G. Akemann and P.H. Damgaard, hep-th/9801133.
\bibitem{GaL} J. Gasser and H. Leutwyler, Ann. Phys. {\bf 158}, 142
(1984); J. Gasser and H. Leutwyler, Nucl. Phys. B{\bf 250}, 465 (1985).

\bibitem{Sharpe} S. Sharpe, Phys. Rev. D 56 (1997) 7052.
\bibitem{qChPT} C. Bernard and M. Golterman, Phys. Rev. D46 (1992)
853; M. Golterman, hep-lat/9411005, {\it Chiral Perturbation Theory
and the quenched approximation of QCD}; S. Sharpe, Phys. Rev. D46
(1992) 3146.        
\bibitem{ColPal} G. Colangelo and E. Pallante, Nucl. Phys. {\bf B520} (1998) 
433.

\bibitem{VWZrep}
J. Verbaarschot, H. Weidenm{\"u}ller, and M. Zirnbauer, Phys. Rep.
{\bf 129} (1985) 367.



\bibitem{zirnall}
M. Zirnbauer, J. Math. Phys. {\bf 37} (1996) 4986.
\bibitem{damgandus}P. Damgaard, J. Osborn, D. Toublan and J. Verbaarschot,
in preparation.

\bibitem{Guhr91}T. Guhr, J. Math. Phys. {\bf 32} (1991) 336.
\bibitem{Alfaro}J. Alfaro, R. Medina and L.F. Urrutia, J. Math. Phys.
{\bf 36} (1995) 3085.
\bibitem{GW}
T. Guhr and T. Wettig, J. Math. Phys. {\bf 37} (1996) 6395.
                                                                                                                                  
\bibitem{Efetov}K.B. Efetov, Adv. Phys. {\bf 32} (1983) 53.
\bibitem{Wegner}F. Wegner, private communication, 1983.
\bibitem{zirnander}M.R. Zirnbauer, Nucl. Phys. {\bf B 265} [FS15] (1986) 375.
\bibitem{Rothstein}
M. J. Rothstein, Trans. Am. Math. Soc. {\bf 299}, 387 (1987).

\bibitem{VZ}J. Verbaarschot and I. Zahed, Phys. Rev. Lett. {\bf 70},
3852 (1993).

\bibitem{HasLeu} P. Hasenfratz and H. Leutwyler, Nucl. Phys. B343
(1990) 241.
\bibitem{Shifman} M. Shifman, A. Vainshtein and V. Zakharov, Nucl. Phys.
{\bf B166} (1980) 493.


\bibitem{Vafa}C. Vafa and E. Witten, Nucl. Phys. {\bf B234} (1984) 173.


                           
                                                                                                        

\end{thebibliography}
\end{document}